\documentclass[preprint,psfig]{aastex6}
\pdfoutput=1
\usepackage{amssymb}
\usepackage{graphicx}
\usepackage{booktabs}
\usepackage{multirow}

\newcommand{\ebv}{$E(B-V)$}

\newcommand{\ha}{\hbox{H$\alpha$}}
\newcommand{\hb}{\hbox{H$\beta$}}
\newcommand{\hg}{\hbox{H$\gamma$}}
\newcommand{\hd}{\hbox{H$\delta$}}

\newcommand{\hii}{\hbox{H\,{\sc ii}}}

\newcommand{\gsim}{\lower.5ex\hbox{$\; \buildrel > \over \sim \;$}}
\newcommand{\lsim}{\lower.5ex\hbox{$\; \buildrel < \over \sim \;$}}
\newcommand{\oi}{\hbox{[O\,{\sc i}]}}
\newcommand{\oii}{\hbox{[O\,{\sc ii}]}}
\newcommand{\oiii}{\hbox{[O\,{\sc iii}]}}
\newcommand{\nii}{\hbox{[N\,{\sc ii}]}}
\newcommand{\sii}{\hbox{[S\,{\sc ii}]}}
\newcommand{\siii}{\hbox{[S\,{\sc iii}]}}

\newcommand{\neiii}{\hbox{[Ne\,{\sc iii}]}}
\newcommand{\hei}{\hbox{[He\,{\sc i}]}}
\newcommand{\ariii}{\hbox{[Ar\,{\sc iii}]}}

\newcommand{\te}{$T_{\mathrm{e}}$}
\newcommand{\nne}{$n_{\mathrm{e}}$}

\newcommand{\myemail}{zesenlin@mail.ustc.edu.cn, xkong@ustc.edu.cn}

\shorttitle{Spectroscopy of \hii\ Regions in M33}

\shortauthors{Lin et al.}

\begin{document}


\title{Spectroscopic Observation and Analysis of \hii\ regions in M33 with MMT: Temperatures and Oxygen Abundances}


\author{Zesen Lin\altaffilmark{1,2}, Ning Hu\altaffilmark{1,2}, Xu Kong\altaffilmark{1,2}, Yulong Gao\altaffilmark{1,2}, Hu Zou\altaffilmark{3}, Enci Wang\altaffilmark{1,2}, Fuzhen Cheng\altaffilmark{1,2}, Guanwen Fang\altaffilmark{4}, Lin Lin\altaffilmark{5}, Jing Wang\altaffilmark{6}}

\altaffiltext{1}{CAS Key Laboratory for Research in Galaxies and Cosmology, Department of Astronomy, University of Science and Technology of China, Hefei 230026, China; \myemail}
\altaffiltext{2}{School of Astronomy and Space Science, University of Science and Technology of China, Hefei 230026, China}
\altaffiltext{3}{National Astronomical Observatories, Chinese Academy of Sciences, Beijing 100012, China}
\altaffiltext{4}{Institute for Astronomy and History of Science and Technology, Dali University, Dali 671003, China}
\altaffiltext{5}{Shanghai Astronomical Observatory, Chinese Academy of Science, 80 Nandan Road, Shanghai, 200030, China}
\altaffiltext{6}{CSIRO Astronomy \& Space Science, Australia Telescope National Facility, P.O. Box 76, Epping, NSW 1710, Australia}


\begin{abstract}

The spectra of 413 star-forming (or \hii) regions in M33 (NGC 598) were observed by using the multifiber spectrograph of Hectospec at the 6.5-m Multiple Mirror Telescope (MMT). By using this homogeneous spectra sample, we measured the intensities of emission lines and some physical parameters, such as electron temperatures, electron densities, and metallicities. Oxygen abundances were derived via the direct method (when available) and two empirical strong-line methods, namely, O3N2 and N2. In the high-metallicity end, oxygen abundances derived from O3N2 calibration were higher than those derived from N2 index, indicating an inconsistency between O3N2 and N2 calibrations. We presented a detailed analysis of the spatial distribution of gas-phase oxygen abundances in M33 and confirmed the existence of the axisymmetric global metallicity distribution widely assumed in literature. Local variations were also observed and subsequently associated with spiral structures to provide evidence of radial migration driven by arms. Our O/H gradient fitted out to 1.1 $R_{25}$ resulted in slopes of $-0.17\pm0.03$, $-0.19\pm0.01$, and $-0.16\pm0.17$ dex $R_{25}^{-1}$ utilizing abundances from O3N2, N2 diagnostics, and direct method, respectively. 

\end{abstract}


\keywords{galaxies: abundances --- galaxies: evolution --- galaxies: individual (M33) --- galaxies: ISM --- galaxies: spiral --- \hii\ regions}



\section{INTRODUCTION}
\label{sec:intro}

Extragalactic star-forming regions are excellent for studying star formation processes, evolution of massive stars, and properties of the surrounding interstellar medium. A wealth of information can be obtained through the conduct of spectral analysis on bright emission lines and stellar continuum. Spectral information from extragalactic \hii\ regions has been traditionally obtained from single-aperture or long-slit observations of some bright parts of the regions through different telescopes at different conditions. We proposed the ``Spectroscopic Observations of the \hii\ Regions In Nearby Galaxies (H2ING)'' Project \citep{Kong2014}, which has performed spectroscopic observations on \hii\ regions in 20 nearby large galaxies since 2008 via the 2.16-m telescope of the National Astronomical Observatories of China (NAOC) and the MMT. In each galaxy, spectra of hundreds of \hii\ regions were observed, and a homogeneous spectrum sample was obtained. In combination with corresponding multiwavelength photometric data, such as the ultraviolet (UV) observation from the {\it Galaxy Evolution Explorer} ({\it GALEX}; \citealt{GildePaz2007}), u-band data of the South Galactic Cap U-band Sky Survey (SCUSS; \citealt{Zhou2016}), 15 intermediate-band data of the Beijing-Arizona-Taipei-Connecticut (BATC) Color Survey of the Sky \citep{Fan1996,Zheng1999}, infrared (IR) observations from {\it Spitzer Space Telescope} \citep{Kennicutt2003a} and {\it Herschel Space Observatory} \citep{Kennicutt2011}, these homogeneous spectra can be used to study many physical properties of nearby galaxies; for example, the star formation rate, spatial distribution of metallicity, and stellar population properties.

As the second paper of this project, we report the observations, data reductions, and analyses on the spectra of the 413 \hii\ regions in M33, which is also known as the Triangulum Galaxy and the third largest galaxy of the Local group. The close distance of M33 makes it one of the nearest disk galaxies of the Milky Way. The NASA/IPAC Extragalactic Database (NED) collects 128 published results of this parameter, which ranges from 600 kpc \citep{Humphreys1980} to 3050 kpc \citep{Willick1997} and has a mean of 878 kpc. In this work, we adopt this mean value as the distance to M33, as well as an inclination of 56\arcdeg\ \citep{Zaritsky1989}, a position angle (PA) of 23\arcdeg\ \citep{Zaritsky1989}, and an $R_{25}$ of 28\arcmin.2\ \citep{Vaucouleurs1959}. This galaxy still maintains a certain level of star-forming activity and has a large amount of \hii\ regions. \cite{Boulesteix1974} compiled a catalog of 369 \hii\ regions based on a complete \ha\ narrow-band photograph survey of M33. More \hii\ region catalogs of M33 were published by later works \citep{Courtes1987b,Calzetti1995,Wyder1997,Hodge1999}.

Combined with the proximity and moderate inclinationthe of M33, the richness of the \hii\ regions in M33 render it ideal for the study of the spatial distribution of metallicity and chemical evolution of spiral galaxies. Since \cite{Searle1971} presented the first spectroscopic observations of the \hii\ regions of M33, many spectroscopic studies have been performed, and the computation of the radial metallicity gradient across the disk of M33 (e.g., \citealt{Smith1975,Kwitter1981,Vilchez1988,Willner2002}) has been attempted. \cite{Crockett2006} was the first to derive the oxygen abundance gradient using \te-based metallicities of 11 \hii\ regions and obtained a slope of $-0.082$ dex $R_{25}^{-1}$. A much larger sample, which included 61 \hii\ regions with temperature-sensitive \oiii$\lambda$4363 line detection, was observed by \cite{Rosolowsky2008}, resulting in an O/H gradient of $-0.19\pm0.08$ dex $R_{25}^{-1}$. Furthermore, \citet{Magrini2010} used their \hii\ region sample, which was observed via MMT, to study the O/H gradient of M33 and found a slope of $-0.30\pm0.12$ dex $R_{25}^{-1}$. Combining a small sample of inner \hii\ regions with other sources in the literature, \citet{Bresolin2011} presented a comparable oxygen gradient of $-0.29\pm0.07$ dex $R_{25}^{-1}$. In addition to the above gradients of radial oxygen abundance based on the analysis of collisionally excited lines (CELs), \citet{ToribioSanCipriano2016} performed a deep spectrophotometry of 11 \hii\ regions in M33 and obtained a gradient of $-0.33\pm0.13$ dex $R_{25}^{-1}$ using the faint optical recombination lines and $-0.36\pm0.07$ dex $R_{25}^{-1}$ based on CELs.

Using the large homogeneous spectra of \hii\ regions, we examine the chemical and physical properties of \hii\ regions and present a detailed study of the spatial distribution of metallicity across the M33 disk. This paper is organized as follows. In Section \ref{sec:obs_redu}, we describe the observations, data reduction, and emission line measurements. In Section \ref{sec:phycond}, we discuss the emission line diagnostics used in this work and our methods for the determination of the physical conditions of the ionized gas. We used three methods to calculate the oxygen abundances of \hii\ regions and provide a brief introduction in Section \ref{sec:abun}. A detailed discussion on the spatial distribution of the gas-phase chemical abundance in M33 is presented in Section \ref{sec:spatial_dist}. Finally, we summarize our results in Section \ref{sec:sum}.

\section{OBSERVATIONS AND DATA REDUCTION}
\label{sec:obs_redu}

\subsection{Target selection}
\label{subsec:target}

\begin{figure}
\centering
\includegraphics[width=\textwidth]{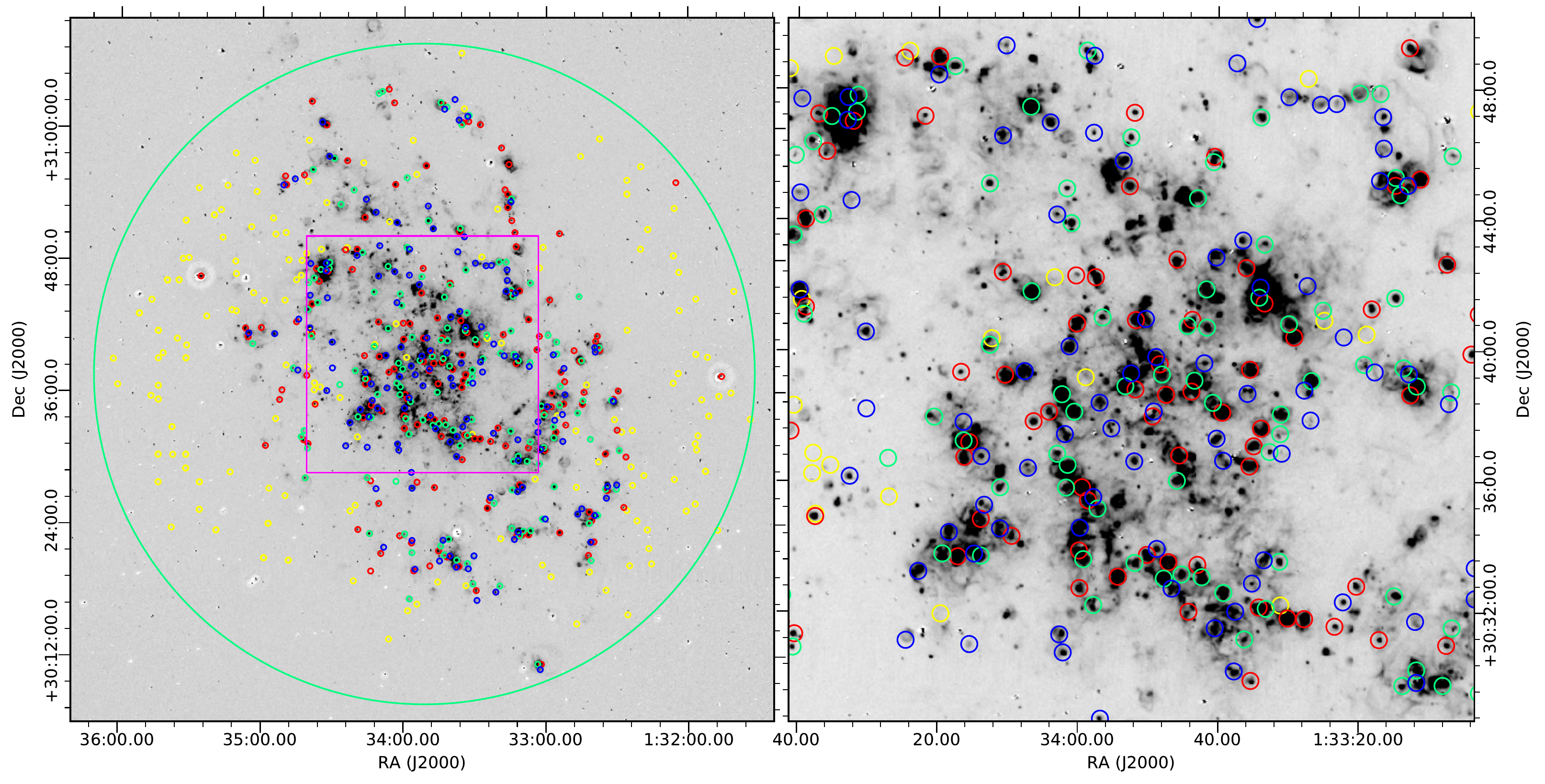}
\caption{Map of the targeted \hii\ regions overlaid on an \ha\ image of M33 (\citealt{Hoopes2001}). The big green circle in the left panel shows the FOV of the Hectospec at the MMT, and the purple square shows the central part of M33 that is enlarged in the right panel. The small red, green and blue circles represent our observed targets on the night of 2013 October 9, 12, and 15, respectively. The small yellow circles are sky-light observations.
\label{fig:FOV}}
\end{figure}

To obtain a \hii\ region sample as complete as possible, we selected the \hii\ region candidates from the continuum-subtracted \ha\ image described in \cite{Hoopes2001}. Candidates were selected with a preference for finding large regions. Using SExtractor \citep{Bertin1996} with irregular filter, we selected candidates around a local maximum of \ha\ flux and required at least 25 pixels (1 pixel $= 2.028$\arcsec) with an \ha\ flux larger than a critical value around it. To eliminate stars from \hii\ region catalog, we used the 2MASS all-sky Point Source Catalog (PSC; \citealt{Cutri2003c}). The Hectospec fiber assignment software xfitfibs\footnote{\url{https://www.cfa.harvard.edu/mmti/hectospec/xfitfibs/}} allows the user to assign rankings to targets. When not prioritized, fiber collisions prevent many objects in the central portion of the center of M33 and thus the objects cannot be observed easily. To reduce the effect of this problem, we assigned priority to \hii\ regions based on their \ha\ fluxes to ensure that bright \hii\ regions in the center would not be missed. Our target catalog contained approximately three times as many targets as the number of fibers available. Therefore, we observed three Hectospec pointings on M33. Given that \hii\ region members are centrally concentrated, approximately 140 fibers were assigned to \hii\ region candidates for each pointing, and the remainder to guide stars or sky. The map of the targeted \hii\ regions, which is overlaid on an \ha\ image of M33, is shown in Figure \ref{fig:FOV}.

\subsection{Observations}
\label{subsec:obs}

The observations discussed in this paper were obtained on the nights of October 9, 12, and 15, 2013, during which the Hectospec fiber-fed spectrograph (\citealt{Fabricant2005}) at the MMT was used. Weather conditions during the night of October 9 were clear, whereas the observations on October 12 and 15 were affected by clouds in some instances. Three 0.5-hour exposures were performed for observations during the first two nights, resulting in three spectra for one target. However, only one 0.5-hour and 0.42-hour exposure were performed on the third night because of the clouds. The instrument deploys 300 fibers over a $1\arcdeg$ diameter field of view, and each fiber has a diameter of 1.5\arcsec\ (corresponding to $\sim6.4$ pc at the distance of M33). We used the 270 line mm$^{-1}$ grating blazed at 5200 \AA\ to provide spectra with a wavelength range of 3650--9200 \AA, a dispersion of 1.2 \AA\ pixel$^{-1}$, and a wavelength resolution of $\sim6$ \AA. 

\subsection{Data reduction}
\label{subsec:redu}

Spectra of the same target were reduced in a uniform manner and combined into one final spectrum. Given that the targets observed on the third night only have two exposures, these two available spectra were used to obtain the final spectrum. We used the processed Hectospec spectra released by the CfA Optical/Infrared Science Archive\footnote{\url{http://oirsa.cfa.harvard.edu/}}. These spectra were homogeneously reduced using the updated version of Hectospec pipeline, HSRED v2.0\footnote{\url{https://www.mmto.org/node/536}}, which was originally written by R. Cool. For spectra observed on the first two nights, the released spectra were similar to those we reduced using the same pipeline. Meanwhile, for those observed on the last night, the released spectra were better than our reduced production, which exhibited some remanent background sky lines. Thus, we used the released version for the subsequent analysis. Two spectra in the released data show significant cutoff in \ha\ lines that might be due to the improper parameter setting in the Hectospec pipeline. We replaced these two spectra by our reduced results. Then, the subsequent flux calibration and de-redshift was performed using the same pipeline.

\begin{figure}
\centering
\includegraphics[scale=0.7, angle=-90]{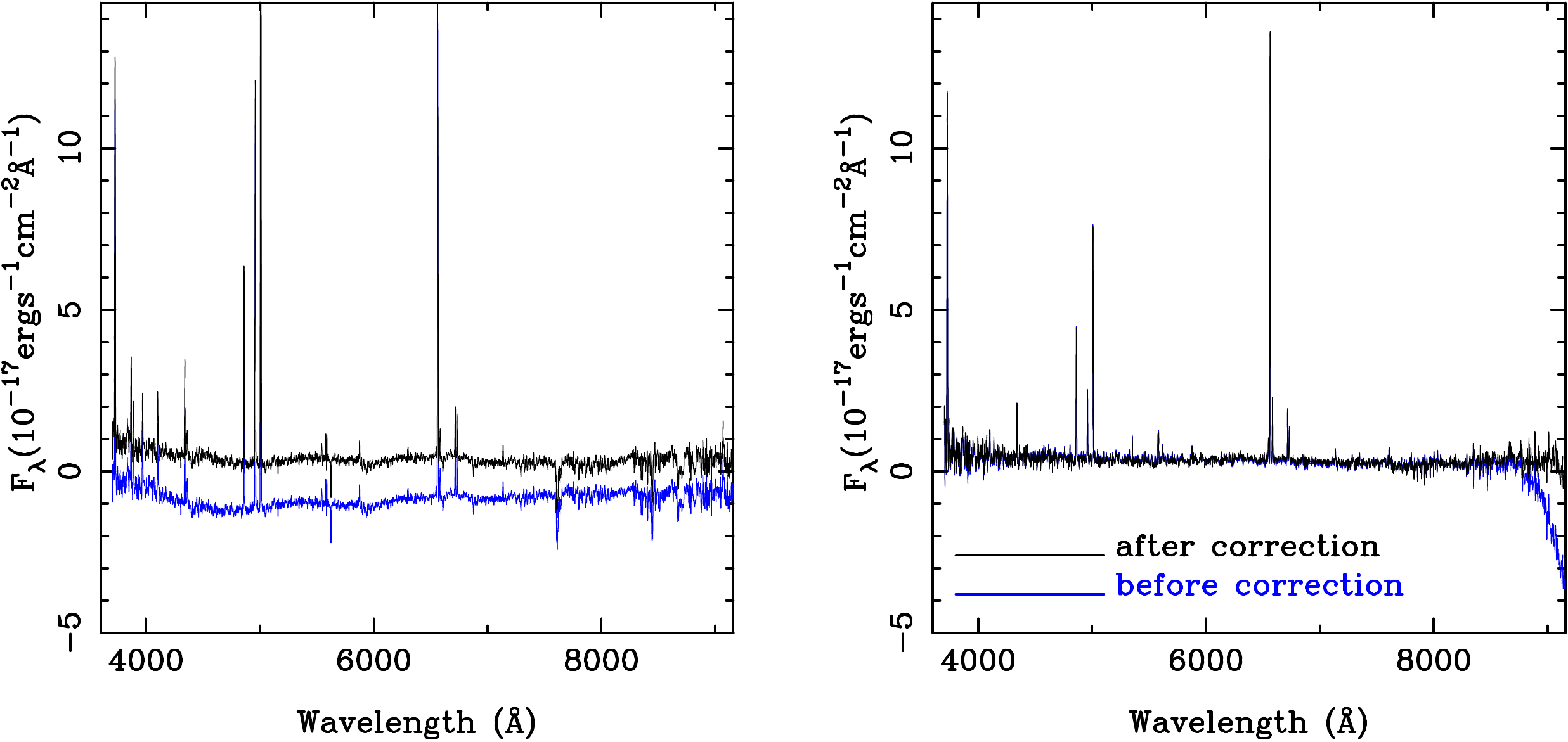}
\caption{Example of flux correction for two types of problematic spectra: negative continuum (left) and decreasing at the red end (right). The blue lines show the spectra before flux correction, and the black lines represent those after correction using BATC photometry. The red lines indicate the zero flux.
\label{fig:fluxcal}}
\end{figure}

After the flux calibration, we found that 45 targets had (1) negative continuum fluxes because of the inaccurate sky subtraction or (2) a sharp increase or decrease at the red end of spectrum. To correct these spectra, we applied an artificial flux correction derived from the BATC image \citep{Ma2001}. We convolved these problematic spectra with the transmission curves of 11 BATC intermediate-band filters, whose center wavelengths ranged from 3890 \AA\ to 9190 \AA. The resulting fluxes were then compared with those obtained from aperture photometry of the BATC map. The correction curves were computed from the differences between the above two set of fluxes via interpolation for the wavelength range covered by the BATC bands and extrapolation for the bluest and reddest end. Figure \ref{fig:fluxcal} presents two typical examples of the problematic spectra and corresponding corrected results.

\begin{figure}
\centering
\includegraphics[scale=0.8]{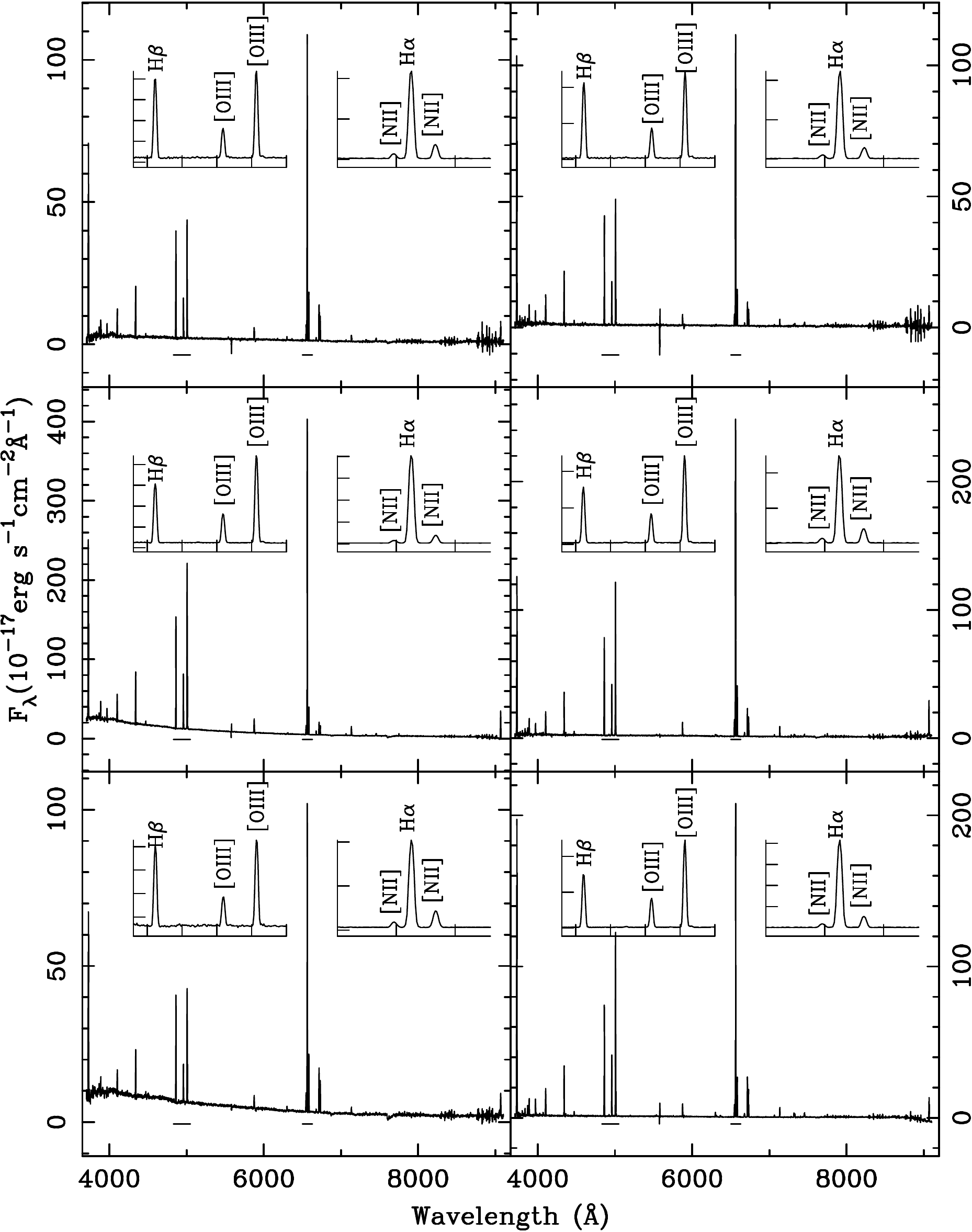}
\caption{Hectospec spectra of six \hii\ regions in M33 for example. Emission lines \ha, \hb, \oiii$\lambda\lambda$4959, 5007, and \nii$\lambda\lambda$6548, 6583 are indicated by underlines and enlarged in two small insets.
\label{fig:spectra}}
\end{figure}

Several spectra were removed because of their significant absorption feature, which suggests that the spectra were dominated by old populations, or extremely low signal-to-noise ratios (S/N) of continua. Finally, we obtained a spectral sample of 413 \hii\ regions\footnote{All spectra and data will be released at \url{http://staff.ustc.edu.cn/~xkong/M33/} after the publication of this work.}. Figure \ref{fig:spectra} shows the Hectospec spectra of six \hii\ region samples of M33 for example. Emission lines \ha, \hb, \oiii$\lambda\lambda$4959, 5007, and \nii$\lambda\lambda$6548, 6583 are indicated by underlines and enlarged into two small insets.

Table \ref{tab:sample} lists all the observed \hii\ regions for which we derived the physical and chemical properties, and information on these \hii\ regions, such as coordinates, identification names, deprojected distances to the center of M33, and PAs, are also presented. The identification names were obtained from the cross-matching among previously published \hii\ region catalogs, such as \citet[BCLMP]{Boulesteix1974}, \citet[VGV86]{Viallefond1986a}, \citet[CPSDP]{Courtes1987b}, \citet[GDK99]{Gordon1999}, and \citet[HBW]{Hodge1999}. To make the identification easier, we used BCLMP as a main catalog for matching because it provided an outline for the description of the relative distribution, as well as the rough shape for every \hii\ region in their catalog. By combining the nearest cross-matching result with that examined macroscopically, we finally found that only 30 \hii\ regions cannot be identified from the catalogs of the previous works. 

\subsection{Emission Line Measurements}
\label{subsec:line_measure}

To obtain reliable emission line fluxes, we first modeled the underlying stellar continuum of each spectrum by using the STARLIGHT\footnote{\url{http://astro.ufsc.br/starlight/}} spectral synthesis code (\citealt{CidFernandes2005}). We fitted the spectrum range of 3700--7500 \AA\ after masking the nebular emission lines, adopting the \cite{Cardelli1989} reddening law and spectrum templates extracted from the \citet[BC03]{Bruzual2003} models. The equivalent widths (EW) of \ha\ absorption lines of the best-fit stellar continua present a median, dispersion, and maximum of $1.9_{-0.6}^{+0.8}$ and 4.5 \AA, while the same parameters are $2.9_{-1.1}^{+1.9}$ and 8.6 \AA\ for \hb\ absorption lines. After subtracting the stellar continuum, each emission line was modeled with one Gaussian profile by using the MPFIT IDL routine (\citealt{Markwardt2009}). The S/N of emission lines were calculated through the method described in \citet{Ly2014}. For \ha\ and \hb\ emission lines, all of our spectra have S/N of greater than 5. The median values and the 16th--84th percentile ranges of the S/N of these two lines are $291_{-153}^{+174}$ and $110_{-80}^{+227}$, respectively. For other emission lines, only those with S/N greater than 3 were considered significantly detected and thus included in the subsequent study.

Given that the intrinsic relative intensities of \ha\ and \hb\ are nearly independent of the physical conditions of the gas, we were able to use the \ha/\hb\ ratios to correct the line of sight reddening. Assuming the intrinsic flux ratio of $\ha/\hb=2.86$ under the Case B recombination, the \cite{Cardelli1989} extinction law with $R_{\mathrm{V}}=3.1$ was employed to obtain the color excesses and de-reddening emission line fluxes. Negative color excesses were all set to be zero. The resulting reddening-corrected emission-line fluxes relative to \hb\ are listed in Table \ref{tab:flux}.

\begin{figure}
\centering
\includegraphics[scale=0.8, angle=-90]{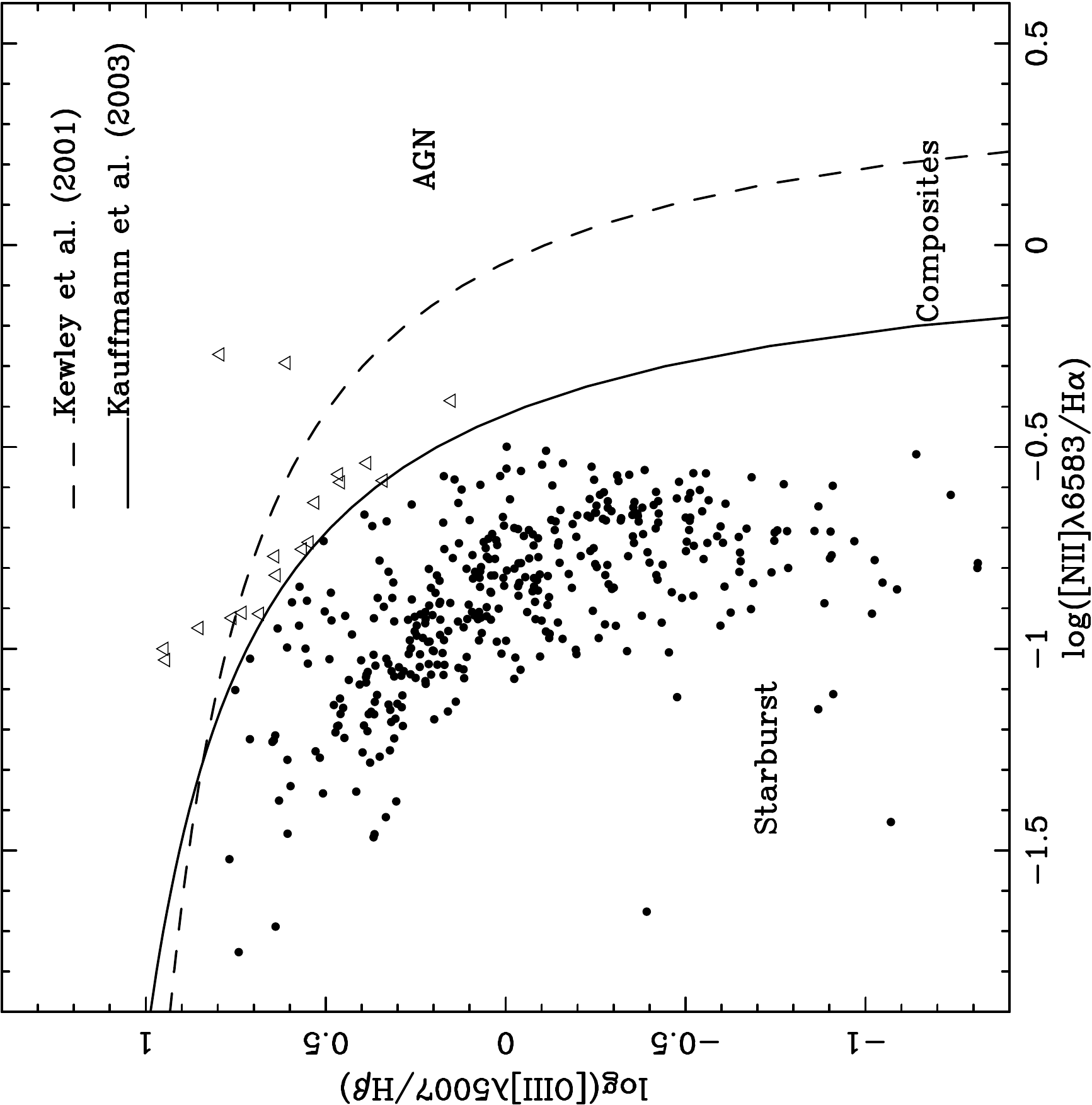}
\caption{BPT diagram exhibiting the excitation properties of our sample. The dotted line represents the theoretical maximum starburst line from \citet{Kewley2001}, whereas the solid line is the boundary between pure star-forming and Seyfert-\hii\ composite region from \citet{Kauffmann2003}. The filled circles indicate sources located in the pure star-forming region. Sources lying above the \citet{Kauffmann2003} boundary, which are likely not star-forming dominated, are marked as open triangles.
\label{fig:bpt}}
\end{figure}

To confirm the \hii\ region nature of our sample, we plotted the Baldwin-Phillips-Terlevich (BPT) diagram \citep{Baldwin1981}, as shown in Figure \ref{fig:bpt}. The boundaries from \citet{Kewley2001} and \citet{Kauffmann2003} were overplotted to divide the $\log(\nii\lambda6583/\ha)-\log(\oiii\lambda5007/\hb)$ plane into three regions: pure star-forming, active galactic nuclei, and composite region, as labeled in Figure \ref{fig:bpt}. Except for 10 \hii\ regions without significant \oiii$\lambda$5007 or \nii$\lambda$6583 detections, most of our regions were located in the pure star-forming region, and only 18 sources lay above the \citet{Kauffmann2003} line, suggesting that the regions are likely not star-forming dominated regions. These exceptions, listed in the footnote of Table \ref{tab:sample}, were excluded from the following analysis. After this selection, we obtained 385 samples of star-forming dominated \hii\ region.

\section{PHYSICAL CONDITIONS OF THE GAS}
\label{sec:phycond}

Physical conditions of the ionized gas in \hii\ regions can be described by several parameters derived from the emission line fluxes. In this study, we used the PyNeb\footnote{\url{http://www.iac.es/proyecto/PyNeb/}} package (\citealt{Luridiana2015}) to calculate electron densities (\nne) and electron temperatures (\te). PyNeb is a python package for analyzing emission lines and includes the Fortran code FIVEL (\citealt{DeRobertis1987}) and the IRAF Nebular package (\citealt{Shaw1995}). We adopted the default atomic data in PyNeb, that is, the transition probabilities of \citet{Wiese1996} and \citet{Storey2000} for O$^{2+}$, \citet{Wiese1996} and \citet{Galavis1997} for N$^+$, \citet{Podobedova2009} for S$^+$, the collision strengths of \citet{Aggarwal1999} for O$^{2+}$, \citet{Tayal2011} for N$^+$, \citet{Tayal2010} for S$^+$. 

Electron temperatures can be determined from some temperature-sensitive auroral lines (e.g., \oiii$\lambda 4363$, \nii$\lambda 5755$, \siii$\lambda 6312$), whereas electron densities can be derived from the ratio between the \sii$\lambda\lambda$6717, 6731 doublet. However, \te\ from \siii$\lambda 6312$ requires detection of \siii$\lambda\lambda$9069, 9532, which is out of our spectrum range. Thus, we measured electron temperatures on the basis of \oiii$\lambda 4363$ and \nii$\lambda 5755$. To draw a reliable conclusion from the following work, we examined each spectrum macroscopically to identify \hii\ regions with not only significant auroral line detection (S/N$\geq$3) but also with well-defined line profile. Finally, we obtained \oiii$\lambda 4363$ and \nii$\lambda 5755$ detections that satisfied the above criteria in the 45 and 27 \hii\ regions, respectively, as indicated in Table \ref{tab:sample}. Among them, 40 and 24 sources were star-forming dominated regions and lay below the \citet{Kauffmann2003} boundary in the BPT diagram.

To obtain a self-consistent estimate of electron densities and electron temperatures, we calculated these two parameters simultaneously using the line flux ratios of \sii$\lambda 6717/\lambda 6731$ and \oiii$\lambda 4363/(\lambda 4959 + \lambda5007)$. This calculation is enable by the PyNeb package. The temperatures derived here were denoted as $T\oiii$. Similarly, for \hii\ regions with both \nii$\lambda 5755$ and \sii$\lambda\lambda$6717, 6731 detections, we calculated electron densities and electron temperatures, denoted as $T\nii$, simultaneously. When \sii$\lambda\lambda$6717, 6731 was undetected, electron temperatures $T\oiii$ and $T\nii$ were calculated under the assumption of $n_\mathrm{e}=100\ \mathrm{cm^{-3}}$. We compared the electron densities obtained from the above two ways but found no significant differences. The former \nne\ was used to calculate the direct metallicity described in the next section because of its large number. When only \sii$\lambda\lambda$6717, 6731 was available and no temperature-sensitive auroral line was present, the electron densities were determined under the assumption that the electron temperature was $10^4$ K.

We repeated the electron temperature and density calculation 2000 times to estimate the uncertainty. In each calculation, flux of each emission line was generated from a Gaussian distribution, assuming the measured flux and uncertainty as the mean and sigma of the distribution, respectively. We used the value derived directly from the measurement fluxes as the final electron temperature or density, and the 68\% range around the measured value in the distribution of 2000 calculations as the corresponding uncertainty. The final electron temperatures and densities are shown in Table \ref{tab:phc}\footnote{One \hii\ region (ID=31) in the \oiii$\lambda$4363 detected subsample gives invalid \te\ value due to its too large $\oiii\lambda4363/\lambda5007$ ratio, whereas three regions (ID=31, 41, 58) in the \nii$\lambda$5755 detected subsample exhibit too large $\nii\lambda5755/\lambda6583$ ratios and suffer the same problem.}. The medians and 16th--84th percentile ranges of these two \te\ were $T\oiii=10100_{-1692}^{+1584}$ K and $T\nii=8900_{-860}^{+2520}$ K. For electron density \nne, we obtained a median and a 68\% scatter of $45_{-30}^{+84}\ \mathrm{cm^{-3}}$.

\begin{figure}
\centering
\includegraphics[scale=0.8, angle=-90]{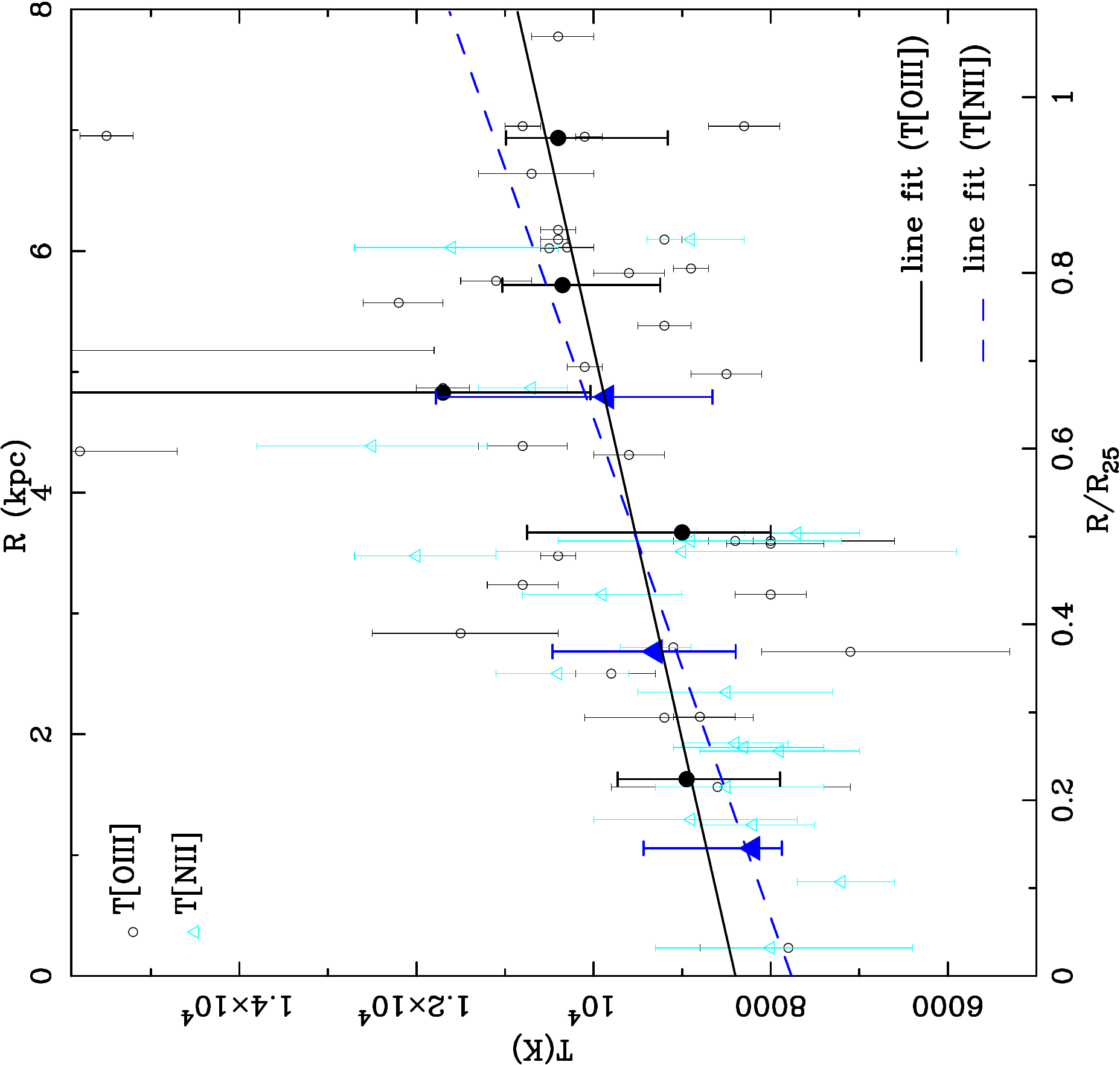}
\caption{Radial distribution of electron temperatures in M33. The grey open circles and cyan open triangles with error bars show the measured values and the uncertainties for $T\oiii$ and $T\nii$, respectively. Data points are divided into radial bins, and the numbers of regions in each bin are the same. The solid symbols are the corresponding binned medians and the 68\% dispersions of the binned distributions. The weighted linear fitting results of the medians are indicated by the black solid line and blue dashed line for $T\oiii$ and $T\nii$, respectively.
\label{fig:te_dist}}
\end{figure}

In Figure \ref{fig:te_dist}, we presented the radial distribution of the 39 (21) \hii\ regions with valid $T\oiii$ ($T\nii$) calculation. Data points were divided into bins of radius, and the numbers of regions in each bin were adjusted to have the same value. Given the small number of \hii\ regions with $T\nii$ measurements, we only set three bins for the $T\nii$ distribution. The medians and the dispersions of the binned distributions were computed and used to perform a weighted linear fitting. The resulting \te\ gradients were as follows:
\begin{eqnarray}
T\oiii/\mathrm{K} = 8398(\pm470)+2243(\pm669)R/R_{25};\\
T\nii/\mathrm{K} = 7765(\pm316)+3520(\pm936)R/R_{25}.
\end{eqnarray}
Our slope of $T\oiii$ was in agreement with \citet{Bresolin2010} in which authors found a gradient of $300\pm60$ K kpc$^{-1}$ at an assumed distance of 840 kpc (i.e., $2067\pm413$ K $R_{25}^{-1}$). The slope of $T\nii$ was slightly steeper than that of $T\oiii$.

The structure of an \hii\ region can be described by a three-zone model (\citealt[hereafter G92]{Garnett1992}) in which every zone can be characterized according to temperature. When this model was used in our work, $T\oiii$ represented the temperature of the high-ionization zone, whereas $T\nii$ depicted the temperature of the low-ionization zone. The temperature of the intermediate-ionization zone can be described by $T\siii$, which was unavailable in our work. Using photoionization models, G92 provided a relation between these temperatures
\begin{equation}
T\nii=T\oii=0.7\times T\oiii+3000\ \mathrm{K}.
\end{equation}
The validity of this relation has been studied by some previous works. \citet{Kennicutt2003} and \citet{Berg2015} used spectral observations of \hii\ regions in M101 and NGC 628, respectively, to examine this relation. However, in both studies, no reliable conclusion was obtained because of their small samples. Furthermore, \citet{Esteban2009} and \cite{Bresolin2009} both found a clear linear relation, which agreed well with the G92 relation, between $T\oiii$ and $T\nii$. 

\begin{figure}
\centering
\includegraphics[scale=0.8, angle=-90]{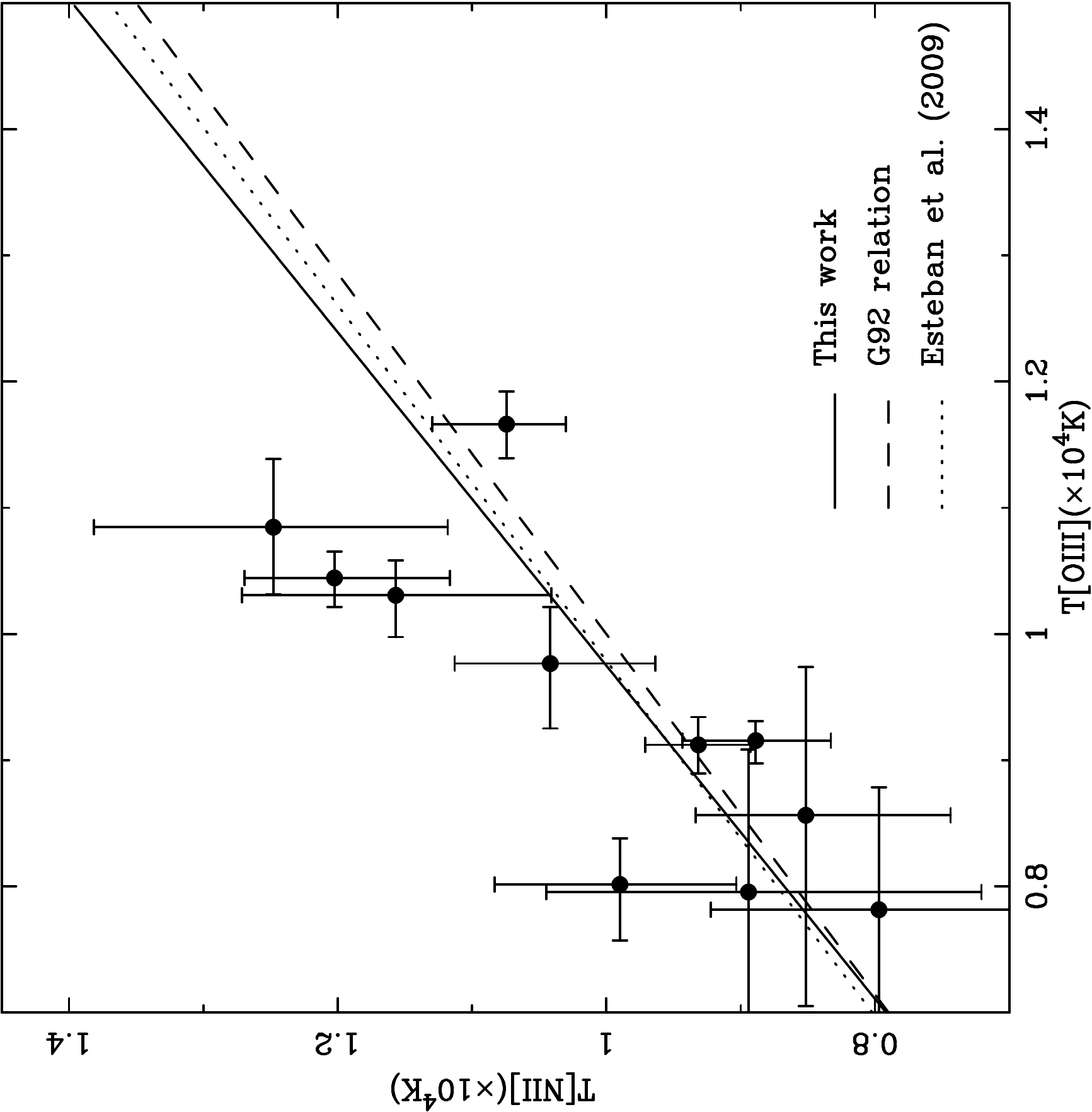}
\caption{Distribution of 11 \hii\ regions with both significant \oiii$\lambda$4363 and \nii$\lambda$5755 detections in the $T\oiii$-$T\nii$ plane. The circles with error bars show the measured values and corresponding uncertainties. A linear fitting result of our data points is indicated by the solid line. The dashed line and the dotted line represent the G92 relation and the relation of \citet{Esteban2009}, respectively.
\label{fig:tem}}
\end{figure}

Using the temperature measurements from \oiii$\lambda$4363 and \nii$\lambda$5755 lines, we examined whether our obtained $T\oiii$ and $T\nii$ follow this relation. Given that the $T\oiii$ versus $T\nii$ relationship is the most difficult to study \citep{Esteban2009}, we only had 11 \hii\ regions that have both valid $T\oiii$ and $T\nii$ measurements. \hii\ regions having auroral line detection flag of 1 (see Table \ref{tab:sample}) or lying outside the \citet{Kauffmann2003} line were not included in this analysis. Figure \ref{fig:tem} shows the locations of these 11 regions in the $T\oiii$--$T\nii$ plane, as well as the G92 relation and the relation derived by \citet{Esteban2009}. Although this subsample was too small to draw a robust conclusion, we still found that all of our \hii\ regions approximately follow the G92 relation. An orthogonal distance regression -- fitting using both uncertainties of $T\oiii$ and $T\nii$ -- was employed and thus a slightly steep slope of 0.76 was obtained.

\section{THE CHEMICAL ABUNDANCES}
\label{sec:abun}

With the measurements of electron temperatures and densities, elemental abundances can be determined from the emission line. This procedure is called the direct method or the \te\ method (\citealt{Kewley2008}). However, in most of our \hii\ regions, the auroral lines \oiii$\lambda$4363 and \nii$\lambda$5755 were extremely weak or undetected. In this case, we used the strong-line method to obtain the metallicities.

\subsection{Direct method}
\label{subsec:te_method}

We used the PyNeb package to calculate the ionic abundance. The O$^+$ and O$^{2+}$ ion abundances were determined from the line intensities of \oii$\lambda$3727\footnote{At the resolution of Hectospec, the \oii$\lambda\lambda$3726, 3728 doublet is unresolved.} and \oiii$\lambda 4959+\lambda 5007$. The electron temperatures of \oii\ were derived from $T\oiii$ using the G92 relation. For \hii\ regions with \sii\ line ratios within the low-density limit (i.e., $n_{\mathrm{e}}<100\ \mathrm{cm}^{-3}$), abundance calculation was assumed to be $n_{\mathrm{e}}=100\pm100\ \mathrm{cm}^{-3}$. The total oxygen abundance was the sum of O$^+$ and O$^{2+}$ ion abundances. The N$^+$/H$^+$ ratios were calculated from the \nii$\lambda 6548+\lambda 6583$ fluxes. We applied a widely used assumption of $\mathrm{N/O}=\mathrm{N}^+/\mathrm{O}^+$ (\citealt{Peimbert1967}) when estimating the abundance of nitrogen. Similar to the calculation process of \te\ and \nne, the uncertainties of elemental abundances were generated from 2000 Monte Carlo simulations considering the uncertainties of line fluxes, \te, and \nne. We did not introduce any uncertainties to represent the systematic errors of the \te\ method. 

Among the 39 \hii\ region with $T\oiii$ calculation, 38 returned valid O$^{2+}$ ion abundances and oxygen abundances. The remaining region, with an ID of 145, showed an extremely large $T\oiii$ value that cannot be processed by the PyNeb code. The N/H abundance was calculated on the basis of the equivalent assumption between $\mathrm{N/O}$ and $\mathrm{N}^+/\mathrm{O}^+$. Thus, the calculation required not only \nii$\lambda$5755 detection but also \oiii$\lambda$4363 detection when $T\oii$ is derived from the G92 relation. As a result, we only had 11 valid measurements of nitrogen abundances from \te\ method. This number is the same as the number of regions with both $T\oiii$ and $T\nii$ measurement.

\subsection{Strong-line methods}
\label{subsec:sl_method}

Although the direct method is the best method to derive gas-phase oxygen abundances, most of our \hii\ regions had no significant auroral line detection. Therefore, we employed two strong-line diagnostics to determine oxygen abundances. The O3N2 index, first introduced by \citet{Alloin1979}, is defined as
\begin{equation}
\mathrm{O3N2}\equiv\log\left(\frac{\oiii\lambda5007}{\hb}\times\frac{\ha}{\nii\lambda6583}\right).
\end{equation}
\citet{Pettini2004} provided an empirical calibration of O3N2 diagnostic that is widely used in literature. Recently, \citet{Marino2013} improved this calibration by using a large \hii\ region sample with \te-determined oxygen abundances. In this work, we used the new calibration from \citet{Marino2013}, which is represented by 
\begin{equation}
12+\log(\mathrm{O/H})=8.533-0.214\times\mathrm{O3N2},
\end{equation}
to calculate the O3N2 abundances with O3N2 ranging from $-1.1$ to 1.7.

Another strong-line index that we used in this work is N2, which is defined as $\mathrm{N2}\equiv\log(\nii\lambda6583/\ha)$. Earlier works, such as that of \citet{Storchi-Bergmann1994} and \cite{Raimann2000}, have reported that the N2 index has a good correlation with oxygen abundance. Thus, it can be used as an abundance estimator. Exploiting their large datasets of \hii\ regions with \te-based abundances, \citet{Marino2013} also presented a calibration for N2 index between $-1.6$ and $-0.2$, which is given by
\begin{equation}
12+\log(\mathrm{O/H})=8.743+0.462\times\mathrm{N2}.
\end{equation}
We used this equation to calculate our N2 abundances. The results and corresponding uncertainties of oxygen abundances derived from these strong-line methods were produced following the method used in the calculation of \te. To account for the uncertainties from the calibrations of \citet{Marino2013}, we assigned uncertainties of 0.18 and 0.16 dex for the O3N2 and N2 calibrations, respectively, which are the rms values of these calibrations. The uncertainties from these calibrations dominated the final errors, whereas the statistical uncertainties exhibited median values of 0.004 and 0.006 dex for O3N2 and N2 calibrations, respectively.

\begin{figure}
\centering
\includegraphics[scale=0.8, angle=-90]{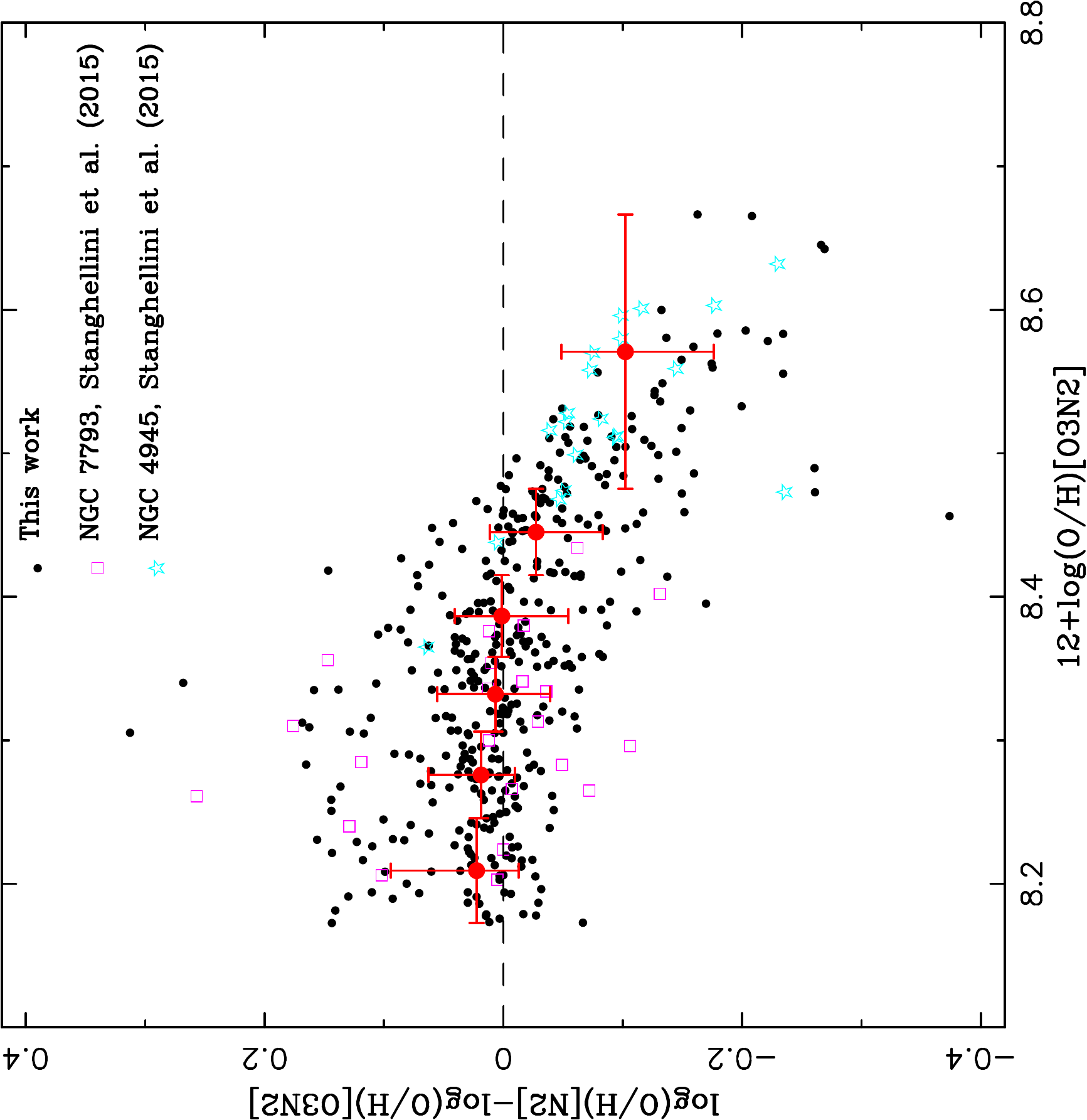}
\caption{Differences of oxygen abundances derived from O3N2 and N2 diagnostics using \citet{Marino2013} calibrations as a function of O3N2-based abundances. The black circles are results of this work. The violet squares and the cyan stars show \hii\ regions from \citet{Stanghellini2015} for NGC 7793 and NGC 4945, respectively. The red circles with error bars are the median values, while the horizontal bars indicate the range of each bin, and the vertical bars represent the 16th--84th percentile ranges of the binned distributions of our data. These bins are adjusted to equalize the number of regions in each bin. 
\label{fig:slm}}
\end{figure}

Our results with regard to metallicities, as well as electron temperatures and densities, are shown in Table \ref{tab:phc}. Figure \ref{fig:slm} presents the differences between oxygen abundances derived from O3N2 and those from N2 index through \citet{Marino2013} calibrations. We also separated the abundances from O3N2 index into six bins and calculated the median and 16th--84th percentile range for each binned distribution, which is shown in figure. For \hii\ regions with $12+\log\mathrm{(O/H)[O3N2]}<8.4$, the oxygen abundances derived from N2 index exhibited nearly the same results. However, when $12+\log\mathrm{(O/H)[O3N2]}>8.4$, the medians in Figure \ref{fig:slm} turned down, $\log\mathrm{(O/H)[N2]}$ tended to be smaller than $\log\mathrm{(O/H)[O3N2]}$, and their differences with oxygen abundances increased. For comparison, we overplotted data of \hii\ regions in NGC 7793 and NGC 4945 from \citet{Stanghellini2015}. Oxygen abundances of \hii\ regions in these two galaxies follow the same trend as our sample. Although these two diagnostics used the same dataset of \hii\ regions for calibration, a discrepancy was detected, indicating inconsistency between O3N2 and N2 calibrations in high-metallicity region. Thus, more data are needed to explore the possible reason of this discrepancy and determine which index has superior performance in this regime.

\section{SPATIAL DISTRIBUTION OF METALLICITY}
\label{sec:spatial_dist}

\subsection{Azimuthal distribution of metals}
\label{subsec:azim_dist}

In the study of how metallicity is distributed in a disk galaxy, an axisymmetric distribution is typically assumed, and a radial distribution is presented. However, whether this assumption holds true for all disk galaxies remains a problem. Given the nearly uniform spatial distribution of our targeted \hii\ regions, we were able to provide a prior check of this assumption before studying the radial oxygen distribution. 

\begin{figure}
\centering
\includegraphics[width=\textwidth]{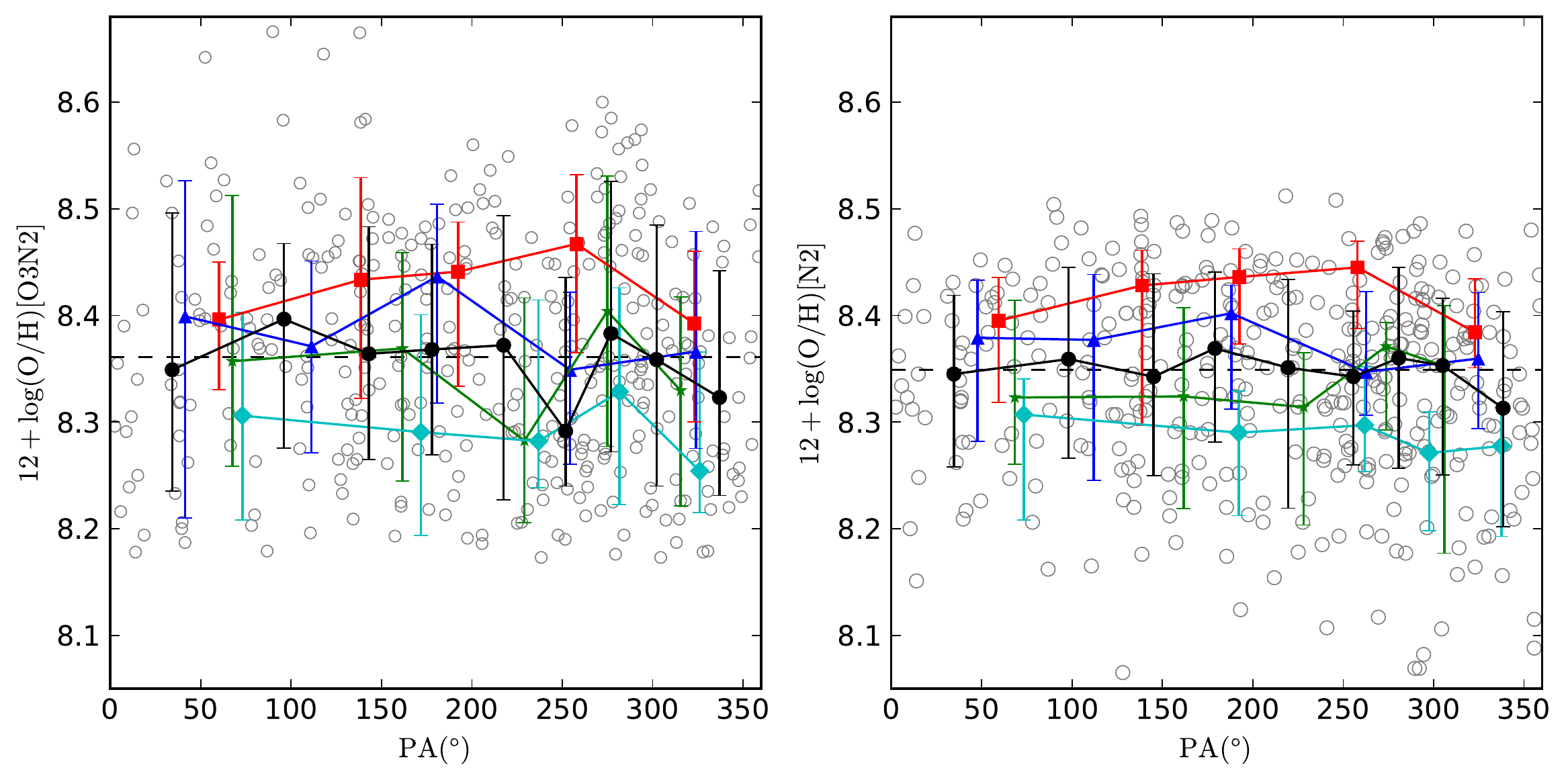}
\caption{Azimuthal oxygen distribution of M33 \hii\ regions. Oxygen abundances were calculated with O3N2 calibration (left) and N2 calibration (right). The position angles are deprojected angles on the disk of M33, measured from north to east (i.e., the north direction is set to be 0\arcdeg, while the PA increases anticlockwise in Figure \ref{fig:FOV}). The grey open circles show the distribution of individual \hii\ region. The black filled circles with the vertical bar connected by lines in each panel indicate the PA-binned medians and corresponding 68\% dispersions for the overall distribution. Data points are also separated into $4\times5$ grids (Figure \ref{fig:grid}) in radius and PA. The median curves are computed for each radial bin and indicated by lines with red squares, blue triangles, green stars, and cyan diamonds from inner to outer radial bins.
\label{fig:az_dist}}
\end{figure}

The azimuthal oxygen distribution of \hii\ regions in M33 is shown in Figure \ref{fig:az_dist}. We divided the sample into nine bins according to their position angles on the galaxy disk (i.e., deprojected position angles) and computed the median value and 16th--84th percentile range of each binned distribution. The numbers of \hii\ regions with oxygen abundance determined from O3N2 and N2 calibrations were 365 and 382, respectively. For each bin, the number of regions was the same. In Figure \ref{fig:az_dist}, black circles with error bars depict the overall azimuthal distribution. Given that a discrepancy occurred between these calibrations (Section \ref{subsec:sl_method}), we only considered features that arise in both distributions.

For O3N2 abundances, the overall median of our sample was 8.36, while all differences between the overall and the binned medians were $\leq 0.04$ dex, except for the sixth ($\sim 225\arcdeg$) PA bins with deviations of $-0.07$ dex. In the case of N2 abundances, our sample presented an overall median of 8.35 and small differences ($\le 0.02$ dex) between the overall and majority of the binned medians, except the last one with the largest difference of $-0.04$. 

\begin{figure}
\centering
\includegraphics[scale=0.7]{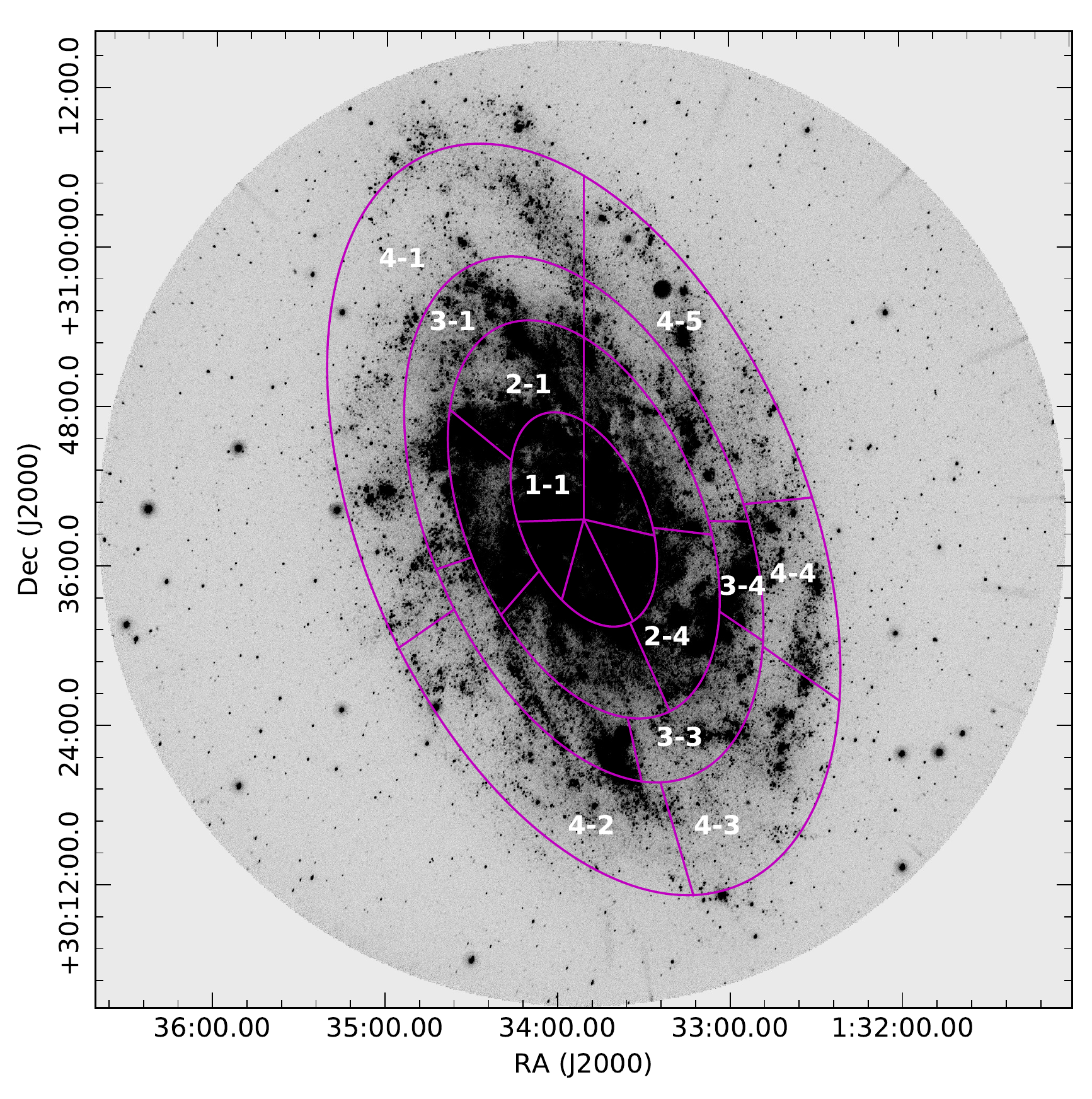}
\caption{Grid of $4\times5$ bins in radius and PA used in the analysis of the azimuthal distribution of the O3N2-based oxygen abundances overlaid on image of NUV from the Nearby Galaxy Survey (NGS; \citealt{GildePaz2007}) of {\it GALEX}. The white text $i-j$ indicates the $j$-th PA bin of the $i$-th radial bin. Only some labels of grids are shown for demonstrating the increasing directions of $i$ and $j$.
\label{fig:grid}}
\end{figure}

Moreover, to search the possible local variation of oxygen abundance, we further divided our sample into $4\times5$ bins according to their located radii and PAs, and ensured that the numbers of \hii\ regions for all grids are constant. Figure \ref{fig:grid} shows this $4\times5$ grid for the O3N2-based oxygen abundances. Hereafter, we used Grid $i-j$ to represent the $j$-th PA bin of the $i$-th radial bin for convenience. For each radical bin, the median curve was calculated and plotted in Figure \ref{fig:az_dist}. Local variations seemed to be significant for O3N2 abundances, and a few of crosses in the median curves of different radial bins were observed. For the four median curves of N2 abundances, a clear radial gradient were detected in nearly all PAs. The first three radial bin showed rapid variations around $200\arcdeg-250\arcdeg$. The most outer bin presented nearly flat median curve, suggesting an axisymmetric distribution of N2 metallicity at that radius.

Combining the results of these two calibrations, we found a significant local variation at a distance of $\sim4.0-5.4$ kpc from the center of M33 and a PA of $\sim200\arcdeg-250\arcdeg$ (i.e., Grid 3-3 and 3-4). After comparing with the \ha\ image \citep{Hoopes2001} and the near-UV (NUV) image from the Nearby Galaxy Survey (NGS; \citealt{GildePaz2007}) of {\it GALEX}, we argue that this abundance jump is due to two loose spiral structures in this region. A spiral arm is present along Grid 2-4 and 3-4. Thus, the median values of these two grids tended to be the same in both azimuthal distributions of O3N2 and N2 abundances. Moreover, a slightly loose arm extends from Grid 3-3 to Grid 4-3 (or even Grid 4-4). The median values of these grids were also nearly the same. The former arm extends from the inner part of M33, where a rich metallicity is retained, whereas the outer arm is relatively metal-poor. The leading side of the metal-poor arm encounters with the trailing edge of the metal-rich arm at the same radius, resulting in the rapid variation in Grid 3-3 and 3-4. This variation implied that the \hii\ regions in the same spiral arm tended to have the same level of metal enrichment, and the metallicity of spiral arm tended to be richer than that of local environment at its trailing edge and poorer at its leading edge in a given radial bin.

\citet{Vogt2017} used deep integral field unit (IFU) observation of the spiral galaxy HCG 91c to study the local variations of the gas-phase oxygen metallicity and found that the metal enrichment prefers to along with the arm of HCG 91c. This conclusion is in agreement with our high-metallicity case (Grid 3-4) but incompatible with our low-metallicity case (Grid 3-3). Simulation results from \citet{Grand2012, Grand2016} demonstrated that the spiral arm can drive a radial migration and lead to systematically more metal-rich (poor) at the trailing (leading) edge of the arm compared with the azimuthal mean metallicity at the same radius. Our results provide observational evidence for this theoretical conclusion, which is also supported by \citet{Sanchez-Menguiano2016}.

Therefore, the oxygen abundances in M33 have a nearly uniform azimuthal distribution within the uncertainty 0.04 dex. \cite{Bresolin2011} presented a similar conclusion for M33 from the small rms scatter (only 0.05 dex) of $R_{23}$-based\footnote{$R_{23}=(\oii\lambda3727+\oiii\lambda\lambda4959,\ 5007)/\hb$.} oxygen abundances without direct azimuthal distribution analysis. By contrast, our large sample allowed us to perform robust direct analysis and provided support for the assumption that the global distribution of oxygen distribution in M33 is axisymmetric. However, local variations also existed and were associated with spiral arm.

\subsection{Radial distribution of oxygen abundances}
\label{subsec:radi_dist}

\begin{figure}
\centering
\includegraphics[scale=0.8, angle=-90]{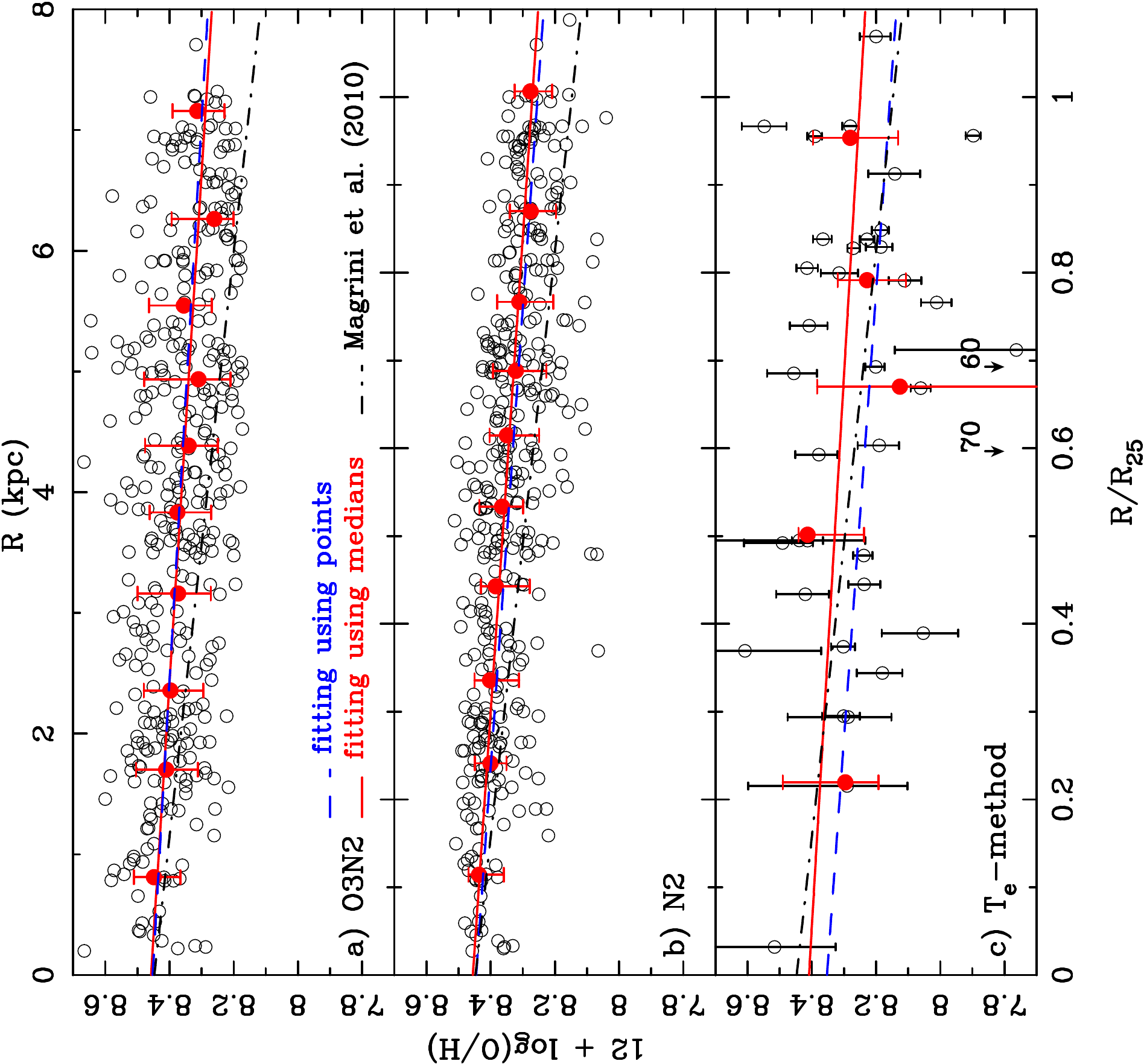}
\caption{Radial oxygen distribution of \hii\ regions in M33. Oxygen abundances were measured from O3N2 calibration (top panel), N2 calibration (middle panel), and \te\ method (bottom panel). The black open circles present the distribution of individual \hii\ region. The red circles with error bars show the binned median and the scatters. The red solid line in each panel represents the best linear fit using the binned medians for each calibration, while the blue dashed line is the best linear fitting with data points. The black dashed-dotted line in each panel is the O/H gradient from \cite{Magrini2010} using their MMT sample only. Two regions with ID of 60 and 70 have \te\ metallicity outside the figure range, and their radial locations are indicated as arrows.
\label{fig:rad_dist}}
\end{figure}

Figure \ref{fig:rad_dist} shows the radial distribution of oxygen abundances. We first performed a linear least-square fit to the relation between the oxygen abundances and deprojected galactocentric distances of \hii\ regions by using the data points directly. The following gradients for oxygen abundances derived from O3N2, N2 calibrations, and \te\ method were then obtained:
\begin{eqnarray}
12+\log\mathrm{(O/H)[O3N2]} = 8.45(\pm0.02)-0.16(\pm0.02)R/R_{25}; \\
12+\log\mathrm{(O/H)[N2]} = 8.44(\pm0.01)-0.19(\pm0.02)R/R_{25}; \\
12+\log\mathrm{(O/H)}[T_{\mathrm{e}}] = 8.35(\pm0.12)-0.20(\pm0.17)R/R_{25}.
\end{eqnarray}
The slope of the gradient derived from the \te\ method was in agreement with that of the N2 index but showed a small intercept. By contrast, the slope from the O3N2-based abundances was slightly flatter than those of the other two. On the one hand, as described in Section \ref{subsec:sl_method}, a discrepancy existed between oxygen abundances derived from these two strong-line calibrations in high-metallicity end (i.e., $12+\log\mathrm{(O/H)[O3N2]}>8.4$). On the other hand, the radial distribution of the \te\ metallicity shows a large dispersion. To reduce possible influence introduced by outliers, we also performed binned analysis for our sample. A weighted linear fitting was employed, resulting in the following:
\begin{eqnarray}
12+\log\mathrm{(O/H)[O3N2]} = 8.46(\pm0.02)-0.17(\pm0.03)R/R_{25}; \\
12+\log\mathrm{(O/H)[N2]} = 8.46(\pm0.01)-0.19(\pm0.01)R/R_{25}; \\
12+\log\mathrm{(O/H)}[T_{\mathrm{e}}] = 8.41(\pm0.11)-0.16(\pm0.17)R/R_{25}.
\end{eqnarray}
The slopes of gradients based on O3N2 and N2 abundances remain nearly unchanged after the binned operation. However, the gradient derived from \te-based metallicity becomes flat although the uncertainty of the slope was large. Overall, the slopes of the radial O/H gradients determined from these three methods were consistent within the uncertainties. Our results were flatter than those of the majority of the previous works listed in Section \ref{sec:intro} but were obtained using a sample that is much larger than the samples used in those works.

\subsection{Two-dimensional distribution of oxygen abundances}
\label{subsec:2d_dist}

\begin{figure}
\includegraphics[width=0.5\textwidth]{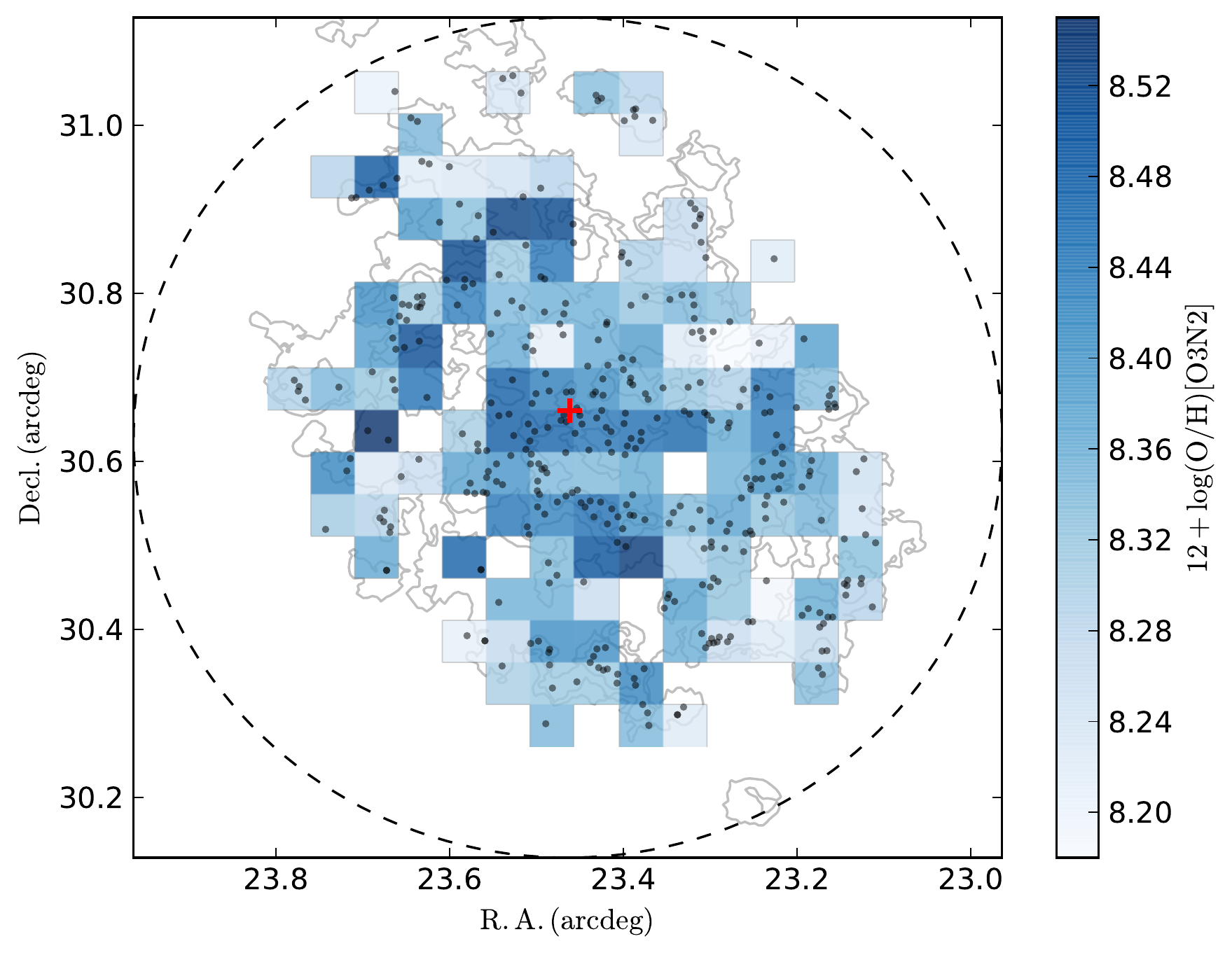}
\includegraphics[width=0.5\textwidth]{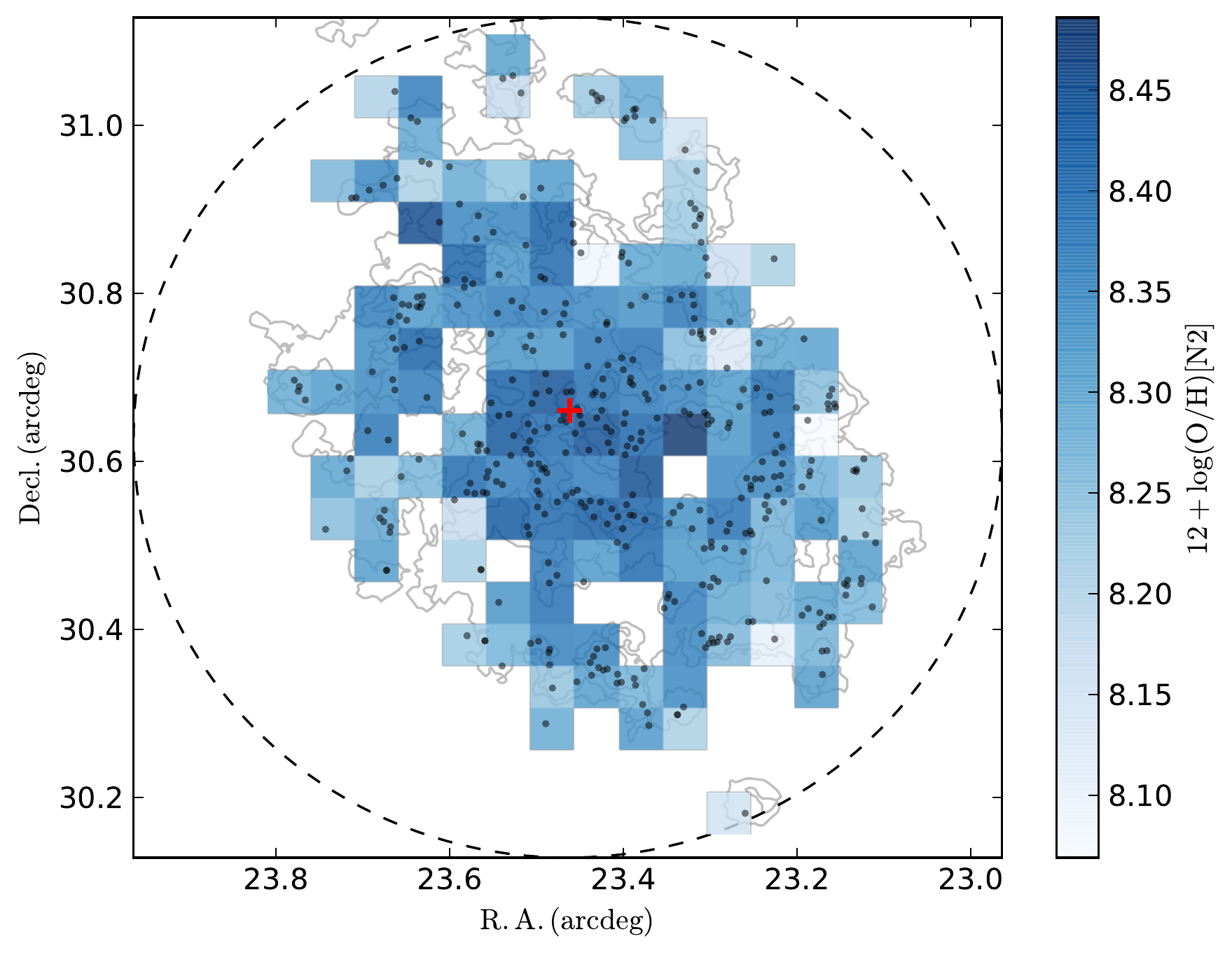}
\caption{Two-dimension distributions of oxygen abundances derived from O3N2 (left panel) and N2 (right panel) calibrations. The small black circles show the distribution of individual \hii\ region. The oxygen abundances of \hii\ regions are averaged in bins of $3\arcmin\times3\arcmin$ (i.e., $0.77\times0.77\ \mathrm{kpc}^2$ in the projected plane) and coded by color. The \ha\ line flux contour of M33 is overplotted by grey lines. The big dashed circles in both panels show the FOV of the Hectospec at MMT, whereas the central red crosses indicate the center of M33 located at R.A. = 1:33:50.89, Decl. = 30:39:36.8 (J2000.0) \citep{Plucinsky2008}. 
\label{fig:2d-abun}}
\end{figure}

Apart from the azimuthal and radial distribution of metals, our large \hii\ region sample enabled us to reconstruct the projected spatial distribution of metals on the disk of M33. In this study, we presented a two-dimensional distribution of oxygen abundances derived from O3N2 and N2 calibrations in Figure \ref{fig:2d-abun}. We divided the projected disk plane into small square bins, each of which had a size of $3\arcmin\times3\arcmin$ (i.e., $0.77\times0.77\ \mathrm{kpc}^2$ in the projected plane). The oxygen abundances of the \hii\ regions falling into the same bin are averaged to obtain the value of each bin. Except for a systematically higher metallicity for O3N2-based distribution relative to N2-based one and a radial gradient, no significant feature can be found in both distributions. 

\citet{Magrini2010} also presented two-dimensional projected distributions of metallicities of \hii\ regions and planetary nebulas (PN) based on a small sample and found the existence of an off-centered metallicity peak located in the southeast direction of M33 in both distributions. Given that the adopted coordinate of the center of M33 and the physical size of grids in the two-dimensional distribution were both nearly the same to ours and those of \citet{Magrini2010}, we were able to conduct a grid-by-grid comparison. The off-centered metallicity peak in the O/H distribution of \hii\ region obtained by \citet{Magrini2010} is located in the second grid below the center of M33. For the same location in our O3N2 and N2 distribution, both grids showed relatively lower metallicity than the center of M33, although only a small number of regions fell into this grid. For the off-centered peak observed in the oxygen abundance of PN by \citet{Magrini2010}, it was located around the third to fourth grid below the center of this galaxy. At the same area, a local high metallicity peak was observed in our O3N2-based distribution but was not more metal-rich than the center of M33. In the case of our N2-based map, this local peak is insignificant. Thus, no such feature was observed in our projected distribution of O3N2 or N2 metallicities. Furthermore, if such off-centered metallicity peak is real, a significant deviation may arise at the corresponding direction (nearly 180\arcdeg\ in our coordinate of azimuthal angle) in our azimuthal distribution of metals in Figure \ref{fig:az_dist}. Specifically, the value of Grid 2-3 should be higher than that of Grid 1-3. However, this phenomenon did not occur, although the O3N2-based metallicity of Grid 2-3 indeed presented a higher value than other grids in the same radial bin. Therefore, both our azimuthal and two-dimensional abundance distributions do not support the existence of this off-centered metallicity maximum. This conclusion was also suggested by \citet{Bresolin2011} through their analysis of radial abundance gradient using \te\ and $R_{23}$-based metallicities.

\section{SUMMARY}
\label{sec:sum}

As one of a series of papers of the H2ING project, this study presented the observations, data reductions, and analyses of spectra in the wavelength range of 3650--9200 \AA\ of 413 \hii\ regions in M33 by using an MMT Hectospec spectrograph. This is by far the largest spectrum sample of \hii\ regions for M33.

Using the \te\ method, we computed the electron temperatures $T\oiii$ and $T\nii$ and the direct oxygen and nitrogen abundances in spectra with temperature-sensitive auroral line detections. We found that the relation between $T\oiii$ and $T\nii$ is consistent with the one presented by G92. We also applied two strong-line diagnostics, namely, O3N2 and N2, to calculate oxygen abundances using calibrations from \cite{Marino2013}. A remarkable discrepancy was detected when $12+\log\mathrm{(O/H)[O3N2]}>8.4$, indicating an inconsistency between O3N2 and N2 calibrations in high-metallicity end.

With this large \hii\ region sample at hand, we presented a detailed study of the spatial distribution of gas-phase oxygen abundances in M33. Our azimuthal metallicity distribution supports the axisymmetric assumption of the global oxygen distribution in M33. However, local variations still existed and were associated with spiral structures. Moreover, we estimated the O/H radial gradient of $-0.17\pm0.03$, $-0.19\pm0.01$, and $-0.16\pm0.17$ dex $R_{25}^{-1}$ making use of O3N2 and N2 diagnostics and \te\ method, respectively. These three slopes agree with each other within the uncertainties but were flatter than most of previous results. From the two-dimensional distribution of oxygen abundances, no significant feature was observed when either O3N2 or N2 calibration was used. Specially, we did not find any evidence indicating the existence of an off-centered metallicity maximum reported by \cite{Magrini2010}.

$\hii$ regions are considered as ideal objects to study the nature of the dispersion of the IRX-$\beta$ relation, i.e., the second parameter of this relation, due to their simple star formation history and strong emission lines (\citealt{Kong2004}). Therefore, combining this large homogeneous spectrum sample of \hii\ regions with multiwavelength photometry data from UV to IR, a subsequent study on the broadening of the IRX-$\beta$ relation will be performed. This study will improve our understanding on the interaction between dust and interstellar radiation field, as well as the formation and evolution of nearby galaxies in a general context.

\acknowledgments
This work uses data obtained through the Telescope Access Program (TAP), which is funded by the National Astronomical Observatories of China, the Chinese Academy of Sciences (the Strategic Priority Research Program, ``The Emergence of Cosmological Structures'' Grant No. XDB09000000), and the Special Fund for Astronomy from the Ministry of Finance. The observations reported in this study were obtained at the MMT Observatory, a joint facility of the Smithsonian Institution and the University of Arizona. This research has made use of the NASA/IPAC Extragalactic Database (NED) which is operated by the Jet Propulsion Laboratory, California Institute of Technology, under contract with the National Aeronautics and Space Administration. This work is supported by the National Basic Research Program of China (973 Program)(2015CB857004), and the National Natural Science Foundation of China (NSFC, Nos. 1320101002, 11433005, 11421303, and 11673004), and the Yunnan Applied Basic Research Projects (2014FB155). EW was supported by the Youth Innovation Fund by University of Science and Technology of China (No. WK2030220019).



\facility{MMT}
\software{SExtractor \citep{Bertin1996}, xfitfibs, HSRED, STARLIGHT \citep{CidFernandes2005}, MPFIT \citep{Markwardt2009}, PyNeb \citep{Luridiana2015}}

\clearpage


\clearpage

\begin{deluxetable}{cccccccccc}
\tabletypesize{\small}
\tabcolsep=4pt
\tablewidth{0pt}
\tablecaption{Sample of \hii\ regions observed in M33.\label{tab:sample}}
\tablehead{
\multirow{2}{*}{ID\tablenotemark{a}} & \multirow{2}{*}{Name\tablenotemark{b}} & \colhead{R.A.} & \colhead{Decl.} & \colhead{$R$\tablenotemark{c}} & \multirow{2}{*}{$R/R_{\mathrm{25}}$\tablenotemark{d}} & \colhead{PA\tablenotemark{e}} & \multicolumn{2}{c}{Auroral Line Detections\tablenotemark{f}} & \multirow{2}{*}{Night\tablenotemark{g}} \\
\cmidrule(lr){8-9}
 & & \colhead{(J2000.0)} &  \colhead{(J2000.0)} & \colhead{(kpc)} &  & \colhead{($\arcdeg$)} & \colhead{\oiii$\lambda$4363} & \colhead{\nii$\lambda$5755} &}
\startdata
1    & \nodata            & 1:32:26.34     & 30:30:10.9      & 7.203 & 1.000 & 273.15  & \nodata & \nodata & 1  \\ 
2    & CPSDP 393       & 1:32:27.24     & 30:25:36.3      & 7.186 & 0.998 & 262.10  & \nodata & \nodata & 1  \\ 
3    & HBW 2           & 1:32:29.11     & 30:30:45.7      & 6.971 & 0.968 & 273.80  & \nodata & \nodata & 2  \\ 
4    & BCLMP 275       & 1:32:29.58     & 30:36:10.6      & 7.208 & 1.001 & 286.45  & 1   & \nodata & 1  \\ 
5    & HBW 3           & 1:32:30.05     & 30:32:37.0      & 6.944 & 0.964 & 278.07  & \nodata & \nodata & 1  \\ 
6    & VGV86 II-001    & 1:32:30.23     & 30:27:38.6      & 6.881 & 0.955 & 265.84  & \nodata & 1   & 3  \\ 
7    & HBW 8           & 1:32:30.37     & 30:27:14.7      & 6.879 & 0.955 & 264.81  & \nodata & \nodata & 2  \\ 
8    & BCLMP 274a      & 1:32:31.60     & 30:35:25.9      & 6.964 & 0.967 & 284.52  & 2   & \nodata & 1  \\ 
9    & BCLMP 274       & 1:32:31.67     & 30:35:15.9      & 6.945 & 0.964 & 284.12  & \nodata & 1   & 3  \\ 
10   & BCLMP 274       & 1:32:32.42     & 30:35:20.1      & 6.881 & 0.955 & 284.19  & 2   & 1   & 2  \\ 

\enddata
\tablecomments{Information of observed \hii\ regions in our sample. The entire table is published in the electronic edition. A portion is shown here for demonstrating its form and content.}
\tablenotetext{a}{IDs of \hii\ regions sorted by right ascension. A total of 10 regions (ID=72, 148, 150, 160, 246, 324, 327, 397, 403, 409) do not have significant \oiii$\lambda$5007 or \nii$\lambda$6583 detections. A total of 18 \hii\ regions with the following IDs lie above the \citet{Kauffmann2003} line at the BPT diagram, indicating that they are likely not star-forming dominated regions: 45, 79, 107, 151, 164, 217, 224, 233, 237, 252, 265, 291, 308, 322, 357, 379, 389, and 398.}
\tablenotetext{b}{The names are from: BCLMP--\citet{Boulesteix1974}; VGV86--\citet{Viallefond1986a}; CPSDP--\citet{Courtes1987b}; GDK99--\citet{Gordon1999}; HBW--\citet{Hodge1999}.}
\tablenotetext{c}{Physical distances to the center of M33 located at R.A. = 1:33:50.89, Decl. = 30:39:36.8 (J2000.0) \citep{Plucinsky2008}. We assumed a distance to M33 of 878 kpc, an inclination of 56\arcdeg\ \citep{Zaritsky1989}, and a position angle of 23\arcdeg\ \citep{Zaritsky1989}.}
\tablenotetext{d}{Assumed an $R_{25}$ of 28\arcmin.2 \citep{Vaucouleurs1959}.}
\tablenotetext{e}{Position angles are deprojected angles on the disk of M33, measured from north to east.}
\tablenotetext{f}{Numbers indicate the detected quality of two auroral Lines, ``1'' represents an emission line with a significant detection (S/N$\geq$3) but without a well-defined profile, whereas ``2'' represents an emission line with not only a significant detection but also a well-defined profile.}
\tablenotetext{g}{Observation nights of each object. ``1,'' ``2,'' and ``3'' indicate the nights of 2013 October 9, 12, and 15, respectively.}
\end{deluxetable}

\clearpage

\begin{splitdeluxetable*}{ccccccccccccBcccccccrrrr}
\tabletypesize{\footnotesize}
\tablecaption{Reddening-corrected emission line fluxes of \hii\ regions in M33. \label{tab:flux}}
\tablehead{
\colhead{ID} & \colhead{\oii} & \colhead{\neiii} & \colhead{\hd} & \colhead{\hg} & \colhead{\oiii} & \colhead{\oiii} & \colhead{\oiii} & \colhead{\nii} & \colhead{\hei} & \colhead{\oi} & \colhead{\nii} & \colhead{\ha} & \colhead{\nii} & \colhead{\sii} & \colhead{\sii} & \colhead{\ariii} & \colhead{\oii} & \colhead{\oii} & \colhead{\hb\tablenotemark{a}} & \colhead{EW(\hb)\tablenotemark{b}} & \colhead{\ebv} & \colhead{Velocity\tablenotemark{c}}\\
 & \colhead{$\lambda$3727} & \colhead{$\lambda$3869} & \colhead{$\lambda$4102} & \colhead{$\lambda$4340} & \colhead{$\lambda$4363} & \colhead{$\lambda$4959} & \colhead{$\lambda$5007} & \colhead{$\lambda$5755} & \colhead{$\lambda$5876} & \colhead{$\lambda$6300} & \colhead{$\lambda$6548} & \colhead{$\lambda$6563} & \colhead{$\lambda$6583} & \colhead{$\lambda$6717} & \colhead{$\lambda$6731} & \colhead{$\lambda$7135} & \colhead{$\lambda$7320} & \colhead{$\lambda$7330} & \colhead{$\lambda$4861} & & &
}
\startdata
1    & 4.065    & \nodata  & 0.531    & 0.608    & \nodata  & 0.100    & 0.301    & \nodata  & \nodata  & 0.293    & 0.117    & 2.474    & 0.335    & 0.564    & 0.370    & \nodata  & \nodata  & \nodata  & 14.9    & 8.8     & \nodata & 56.95    \\
 & 0.095    & \nodata  & 0.054    & 0.043    & \nodata  & 0.009    & 0.027    & \nodata  & \nodata  & 0.017    & 0.005    & 0.050    & 0.014    & 0.017    & 0.014    & \nodata  & \nodata  & \nodata  & 0.5     & & &\\
2    & 3.164    & \nodata  & 0.223    & 0.452    & \nodata  & 0.436    & 1.309    & \nodata  & 0.049    & 0.028    & 0.083    & 2.693    & 0.239    & 0.241    & 0.179    & 0.037    & 0.046    & 0.040    & 155.2   & 59.8    & \nodata & 69.64    \\
 & 0.058    & \nodata  & 0.010    & 0.013    & \nodata  & 0.007    & 0.020    & \nodata  & 0.003    & 0.002    & 0.002    & 0.050    & 0.005    & 0.006    & 0.004    & 0.002    & 0.002    & 0.002    & 3.2     & & &\\
3    & 3.879    & \nodata  & 0.220    & 0.399    & \nodata  & 0.516    & 1.548    & \nodata  & 0.186    & 0.117    & 0.091    & 2.860    & 0.261    & 0.243    & 0.187    & \nodata  & 0.075    & \nodata  & 24.2    & 65.6    & 0.0274  & 42.08    \\
 & 0.069    & \nodata  & 0.010    & 0.012    & \nodata  & 0.008    & 0.023    & \nodata  & 0.006    & 0.005    & 0.002    & 0.053    & 0.007    & 0.006    & 0.005    & \nodata  & 0.004    & \nodata  & 0.5     & & &\\
4    & 4.193    & \nodata  & 0.197    & 0.459    & 0.060    & 0.313    & 0.940    & \nodata  & 0.149    & \nodata  & 0.086    & 2.602    & 0.247    & 0.322    & 0.261    & 0.040    & \nodata  & \nodata  & 57.8    & 40.9    & \nodata & 33.30    \\
 & 0.078    & \nodata  & 0.010    & 0.014    & 0.006    & 0.006    & 0.017    & \nodata  & 0.005    & \nodata  & 0.002    & 0.048    & 0.006    & 0.008    & 0.007    & 0.003    & \nodata  & \nodata  & 1.2     & & &\\
5    & 2.618    & \nodata  & 0.360    & 0.446    & \nodata  & 0.667    & 2.001    & \nodata  & \nodata  & 0.096    & 0.053    & 2.099    & 0.153    & 0.175    & 0.150    & 0.061    & 0.077    & 0.077    & 18.0    & 7.1     & \nodata & 40.41    \\
 & 0.122    & \nodata  & 0.063    & 0.060    & \nodata  & 0.018    & 0.055    & \nodata  & \nodata  & 0.013    & 0.003    & 0.042    & 0.010    & 0.010    & 0.011    & 0.008    & 0.008    & 0.008    & 0.8     & & &\\
6    & 5.019    & 0.893    & \nodata  & 0.565    & \nodata  & 0.413    & 1.233    & 0.104    & 0.223    & 0.760    & 0.149    & 2.860    & 0.427    & 1.259    & 0.950    & \nodata  & \nodata  & 0.220    & 28.9    & 31.2    & 0.1094  & 17.20    \\
 & 0.092    & 0.022    & \nodata  & 0.015    & \nodata  & 0.007    & 0.021    & 0.006    & 0.008    & 0.016    & 0.003    & 0.052    & 0.010    & 0.023    & 0.019    & \nodata  & \nodata  & 0.015    & 0.6     & & &\\
7    & 3.876    & \nodata  & 0.365    & 0.562    & \nodata  & 0.115    & 0.345    & \nodata  & 0.156    & 0.420    & 0.137    & 2.860    & 0.395    & 1.137    & 0.755    & \nodata  & \nodata  & \nodata  & 24.0    & 25.9    & 0.0532  & 51.29    \\
 & 0.076    & \nodata  & 0.020    & 0.020    & \nodata  & 0.004    & 0.013    & \nodata  & 0.008    & 0.012    & 0.003    & 0.053    & 0.009    & 0.024    & 0.017    & \nodata  & \nodata  & \nodata  & 0.6     & & &\\
8    & 1.577    & 0.374    & 0.273    & 0.474    & 0.043    & 1.713    & 5.139    & \nodata  & 0.112    & 0.050    & 0.059    & 2.860    & 0.171    & 0.263    & 0.178    & 0.097    & 0.020    & 0.017    & 245.5   & 223.3   & 0.0026  & 18.16    \\
 & 0.028    & 0.009    & 0.006    & 0.010    & 0.002    & 0.022    & 0.067    & \nodata  & 0.003    & 0.002    & 0.001    & 0.053    & 0.003    & 0.005    & 0.004    & 0.002    & 0.001    & 0.001    & 4.6     & & &\\
9    & 2.735    & 0.257    & 0.237    & 0.456    & \nodata  & 1.182    & 3.544    & 0.007    & 0.114    & 0.056    & 0.092    & 2.860    & 0.263    & 0.319    & 0.226    & 0.134    & 0.033    & 0.033    & 319.9   & 64.4    & 0.0215  & 33.09    \\
 & 0.050    & 0.010    & 0.008    & 0.011    & \nodata  & 0.016    & 0.049    & 0.007    & 0.004    & 0.003    & 0.002    & 0.053    & 0.005    & 0.007    & 0.006    & 0.004    & 0.002    & 0.002    & 6.5     & & &\\
10   & 1.906    & 0.458    & 0.268    & 0.465    & 0.035    & 1.710    & 5.131    & 0.006    & 0.118    & 0.096    & 0.092    & 2.786    & 0.263    & 0.378    & 0.280    & 0.105    & 0.027    & 0.026    & 261.1   & 244.5   & \nodata & 15.49    \\
 & 0.033    & 0.009    & 0.006    & 0.009    & 0.002    & 0.022    & 0.067    & 0.001    & 0.003    & 0.003    & 0.002    & 0.051    & 0.004    & 0.008    & 0.006    & 0.003    & 0.002    & 0.001    & 4.8     & & &\\

\enddata
\tablecomments{Reddening-corrected emission line fluxes of \hii\ regions in M33. Emission line fluxes except for \hb\ are normalized to $F$(\hb). For each \hii\ region, the first line is a measured flux for each emission line, whereas the second line presents the corresponding measurement error. The entire table is published in the electronic edition. A portion is shown here for demonstrating its form and content.}
\tablenotetext{a}{Reddening-corrected \hb\ fluxes have units of $10^{-17}$ erg s$^{-1}$ cm$^{-2}$.}
\tablenotetext{b}{\hb\ emission line equivalent widths in units of \AA\ corrected for stellar absorption. Two \hii\ regions (ID=160, 350) have negative local continuum fluxes, resulting in incorrect EWs of \hb. These EWs are excluded from this table.}
\tablenotetext{c}{Radial velocity for each region, in units of $\mathrm{km\ s^{-1}}$. System velocity of M33, assuming a value of $-179\ \mathrm{km\ s^{-1}}$ \citep{Vaucouleurs1991}, has been subtracted.}
\end{splitdeluxetable*}

\clearpage

\begin{deluxetable}{crrrrrrr}
\tabletypesize{\footnotesize}
\tablewidth{0pt}
\tablecaption{Physical conditions and metallicity of \hii\ regions in M33.\label{tab:phc}}
\tablehead{
\multirow{2}{*}{ID\tablenotemark{a}} & \colhead{$T\oiii$\tablenotemark{b}} & \colhead{$T\nii$\tablenotemark{c}} & \colhead{\nne\tablenotemark{d}} & \multicolumn{3}{c}{12+$\log\mathrm{(O/H)}$} & \multirow{2}{*}{12+$\log\mathrm{(N/H)}$\tablenotemark{g}} \\
\cmidrule(lr){5-7}
    & \colhead{(K)} & \colhead{(K)}  & \colhead{(cm$^{-3}$)} & \colhead{[\te]}  & \colhead{[O3N2]\tablenotemark{e}}  & \colhead{[N2]\tablenotemark{f}}   &
}  
\startdata
1    & \nodata & \nodata & \nodata & \nodata & $8.459 _{-0.010 }^{+0.010 }$ & $8.342 _{-0.009 }^{+0.009 }$ & \nodata \\ 
2    & \nodata & \nodata & $71  _{-34  }^{+35  }$ & \nodata & $8.283 _{-0.003 }^{+0.004 }$ & $8.257 _{-0.006 }^{+0.005 }$ & \nodata \\ 
3    & \nodata & \nodata & $113 _{-40  }^{+42  }$ & \nodata & $8.270 _{-0.004 }^{+0.004 }$ & $8.263 _{-0.006 }^{+0.007 }$ & \nodata \\ 
4    & $32600   _{-3100    }^{+3300    }$ & \nodata & $224 _{-73  }^{+74  }$ & \nodata & $8.320 _{-0.004 }^{+0.004 }$ & $8.271 _{-0.005 }^{+0.006 }$ & \nodata \\ 
5    & \nodata & \nodata & $245 _{-109 }^{+135 }$ & \nodata & $8.225 _{-0.008 }^{+0.008 }$ & $8.218 _{-0.014 }^{+0.013 }$ & \nodata \\ 
6    & \nodata & \nodata & $87  _{-27  }^{+31  }$ & \nodata & $8.337 _{-0.004 }^{+0.004 }$ & $8.361 _{-0.006 }^{+0.006 }$ & \nodata \\ 
7    & \nodata & \nodata & \nodata & \nodata & $8.448 _{-0.005 }^{+0.005 }$ & $8.346 _{-0.006 }^{+0.006 }$ & \nodata \\ 
8    & $10800   _{-200     }^{+200     }$ & \nodata & \nodata & $8.280 _{-0.025 }^{+0.025 }$ & \nodata & $8.177 _{-0.005 }^{+0.005 }$ & \nodata \\ 
9    & \nodata & $13500   _{-6300    }^{+6500    }$ & $24  _{-22  }^{+24  }$ & \nodata & $8.194 _{-0.004 }^{+0.004 }$ & $8.264 _{-0.005 }^{+0.005 }$ & \nodata \\ 
10   & $10100   _{-200     }^{+100     }$ & $11600   _{-1000    }^{+1000    }$ & $70  _{-28  }^{+32  }$ & $8.390 _{-0.022 }^{+0.024 }$ & \nodata & $8.270 _{-0.005 }^{+0.005 }$ & $7.168 _{-0.084 }^{+0.101 }$ \\ 

\enddata
\tablecomments{Electron temperatures, electron densities, and metallicities derived from \te\ method and two strong-line methods of \hii\ regions in M33. These properties of \hii\ regions with auroral line flag of 1 in Table \ref{tab:sample} are also provided. The entire table is published in the electronic edition. A portion is shown here for demonstrating its form and content.}
\tablenotetext{a}{IDs of \hii\ regions sorted by right ascension.}
\tablenotetext{b}{Electron temperatures calculated from \oiii$\lambda$4364 line.}
\tablenotetext{c}{Electron temperatures calculated from \nii$\lambda$5755 line.}
\tablenotetext{d}{Electron densities were computed from an iterative process using \oiii\ and \sii\ line ratios or a single calculation using \sii\ line ratios only but assuming a \te\ of $10^4$ K, see text for detail.}
\tablenotetext{e}{Oxygen abundance derived from O3N2 index using calibration from \cite{Marino2013}. The systematic uncertainty of this calibration is 0.18.}
\tablenotetext{f}{Oxygen abundance derived from N2 index using calibration from \cite{Marino2013}. The systematic uncertainty of this calibration is 0.16.}
\tablenotetext{g}{Nitrogen abundances derived from \te\ method using $T\nii$.}
\end{deluxetable}







\end{document}